\documentclass[11pt,a4paper]{article}
\pdfoutput=1
\usepackage{jheppub}
\usepackage{amsfonts, amssymb, amsmath, amsthm}
\usepackage{mathrsfs}
\usepackage{graphicx}
\usepackage[utf8]{inputenc}
\usepackage[english]{babel}
\usepackage{slashed}
\usepackage{tikz}
\usepackage[indent]{parskip}
\usepackage{color}
\definecolor{dark-gray}{gray}{0.20}
\definecolor{gray}{gray}{0.30}
\definecolor{light-gray}{gray}{0.80}
\definecolor{dark-red}{rgb}{0.7,0,0}
\definecolor{dark-green}{rgb}{0.1,0.4,0}
\definecolor{dark-blue}{rgb}{0.3,0.3,0.7}
\definecolor{light-blue}{rgb}{0.8,0.8,1}
\definecolor{blue}{rgb}{0,0,1}
\definecolor{red}{rgb}{1,0,0}
\definecolor{green}{rgb}{0,1,0}

\usepackage{hyperref}
\hypersetup{
	colorlinks=true,
	linkcolor=dark-blue,
	citecolor=dark-red,
	urlcolor=dark-blue,
	linktoc=page
}

\def\cB{{\cal B}}

\def\cF{{\cal F}}
\def\cI{{\cal I}}

\def\cN{{\cal N}}
\def\cO{{\cal O}}

\def\cC{{\cal C}}

\def\SO{{\rm SO}}

\def\SU{{\rm SU}}

\def\i{{\rm i}}

\theoremstyle{definition}

\theoremstyle{remark}

\newcommand{\be}{\begin{equation}}
\newcommand{\ee}{\end{equation}}
\newcommand{\ba}{\begin{aligned}}
\newcommand{\ea}{\end{aligned}}
\newcommand{\bea}{\begin{eqnarray}}
\newcommand{\eea}{\end{eqnarray}}

\newcommand{\mathd}{\mathrm{d}}
\newcommand{\mathe}{\mathrm{e}}
\newcommand{\mathi}{\mathrm{i}}

\newcommand{\e}{\epsilon}
\newcommand{\lam}{\lambda}
\newcommand{\BV}{\mathbb{V}}

\newcommand{\BC}{\mathbb{C}}
\newcommand{\BR}{\mathbb{R}}

\newcommand{\BZ}{\mathbb{Z}}
\newcommand{\WPL}{\mathbb{WP}^1_{a,b}}
\newcommand{\VVV}{\mathrm{vol}}
\newcommand{\NN}{\bar{\nu}}
\newcommand{\squb}{\mathsf{b}}
\title{M2-brane partition functions and HD supergravity\\ from equivariant volumes}

\author[a]{Luca Cassia}
\author[b,c]{and Kiril Hristov}

\affiliation[a]{School of Mathematics and Statistics, The University of Melbourne Parkville,\\ Melbourne, VIC 3010, Australia}
\affiliation[b]{Faculty of Physics, Sofia University ``St.\ Kliment Ohridski'',\\ J. Bourchier Blvd. 5, 1164 Sofia, Bulgaria}
\affiliation[c]{INRNE, Bulgarian Academy of Sciences, Tsarigradsko Chaussee 72, 1784 Sofia, Bulgaria}

\emailAdd{luca.cassia@unimelb.edu.au}
\emailAdd{khristov@phys.uni-sofia.bg}

\abstract{
\noindent We use the results of our companion paper \cite{Cassia:2025aus}, which explores the equivariant generalization of the constant-map contribution in topological string theory on toric Calabi--Yau manifolds $X$, to establish a holographic correspondence with M2-brane partition functions. We rigorously test this conjecture within the perturbative regime, incorporating all finite $N$ corrections on the field theory side. Our approach involves additional key details, such as incorporating effective 4d higher-derivative supergravity corrections to introduce refinement. A central result is the derivation of the Airy function representation for the squashed $S^3$ partition function of the field theory, for an arbitrary squashing parameter. We demonstrate that this Airy function structure is universal across all M2-brane models and provide a general expression in terms of the equivariant volume of $X$, incorporating the mesonic deformations corresponding to complexified masses. This expression is then evaluated explicitly for several examples, including ABJM theory, its flavored generalizations, circular quivers, and beyond, demonstrating agreement with the available field theory localization results. We extend the analysis to the superconformal and twisted indices of M2-brane models, and their spindle generalizations, leaving their full perturbative completion for future work. Finally, we explore avenues for generalizing these results to other brane systems, explicitly applying the idea to D3-branes.
}
\date{\today}
\begin{document}
\maketitle

\section{Summary of results}

\begin{figure}[ht]
\hspace{-2em}
\begin{tikzpicture}[scale=1, every node/.style={scale=.9}]
	
	\node at (-2.5,0){bulk};
	
	\node at (-2.5,-3.1){bndr.};

	\draw[fill = white,thick, rounded corners, draw = blue] (-1.9,-.75) rectangle (0.7,.75);
	\node at (-0.6,0.4){\large equivariant};
\node at (-0.6,0){\large topological};
\node at (-0.6,-0.4){\large strings, $F^\text{top}_X (\lambda)$};
	
	\draw[fill = white,thick, rounded corners, draw = blue] (2.3,-0.75) rectangle (4.7,.75);
	\node at (3.5,0.4){\large constant};
	\node at (3.5,-0){\large maps terms,};
	\node at (3.5,-0.4){\large $\sum_{\mathfrak{g}} N^X_{\mathfrak{g}, 0} (\lambda) $};

	\draw[fill = white,thick, rounded corners, draw = yellow] (6.5,-.75) rectangle (9.1,.75);
	\node at (7.8,0.4){\large higher der.};
	\node at (7.8,-0){\large sugra on $L$, };
	\node at (7.8,-0.4){\large $I_L^\text{HD} (G_{\rm N})$};

	\draw[fill = white,thick, rounded corners, draw = yellow] (10.85,-.75) rectangle (13.15,.75);
	\node at (12,0.4){\large 2-der.};
	\node at (12,-0){\large sugra on $L$, };
	\node at (12,-0.4){\large $I_L (G_{\rm N})$};

	\draw[->,>=stealth, black, ultra thick] (.8,0) -- (2.2,0);
	\node at (1.5,0.3){perturb.};
		\node at (1.5,-0.3){trunc.};

	\draw[<->,>=stealth, green, ultra thick] (4.8,0) -- (6.4,0);
	\node at (5.6,0.3){saddle pt.};
	\node at (5.6,-0.3){approx.};

	\draw[->,>=stealth, black, ultra thick] (9.2,0) -- (10.8,0);
	\node at (9.95,0.3){$G_{\rm N}^{-1} \rightarrow \infty$};

	\draw[fill = white,  thick, rounded corners, draw = yellow] (-1.9,-3.75) rectangle (.7,-2.25);
	\node at (-0.6,-2.6){\large M2-branes};
	\node at (-0.57,-3.0){\large on tip of $X$,};
\node at (-0.6,-3.4){\large $Z_{\rm M2} (N_{\rm M2})$};	

	\draw[fill = white,  thick, rounded corners, draw = yellow] (2.3,-3.75) rectangle (4.7,-2.25);
	\node at (3.5,-3){\Large $Z^\text{pert}_{\rm M2} (N_{\rm M2})$};

	\draw[fill = white, thick, rounded corners, draw = yellow] (10.85,-3.75) rectangle (13.15,-2.25);
	\node at (12,-2.7){\large leading free};
	\node at (12,-3.2){\large energy, $\cF (N)$};	

	\draw[<->,>=stealth, black, ultra thick] (12,-0.9) -- (12,-2.1);

	\draw[<->,>=stealth, green, ultra thick] (3.5,-0.9) -- (3.5,-2.1);

	\draw[<->,>=stealth, green, ultra thick] (-0.5,-0.9) -- (-0.5,-2.1);
	\node at (-2.1,-1.5){AdS/CFT};

	\draw[->,>=stealth, black, ultra thick] (.8,-3) -- (2.2,-3);	
		\node at (1.5,-2.7){perturb.};
		\node at (1.5,-3.3){trunc.};

	\draw[->,>=stealth, black, ultra thick] (5,-3) -- (10.5,-3);
	\node at (7.7,-2.7){$N \rightarrow \infty$};

	\draw[->,>=stealth, gray, thick] (5,1.1) -- (4.3,.84);
	\node at (5.5,1.1){in \cite{Cassia:2025aus}};	

	\draw[->,>=stealth, red, thick] (5.7,-1.4) -- (5.6,-0.6);
	\draw[->,>=stealth, red, thick] (5.1,-1.7) -- (3.9,-1.5);
	\node at (5.7,-1.7){{\color{red} here}};

	\end{tikzpicture}
	\caption{Schematic diagram of the proposed relations,  suppressing additional indices and equivariant parameters. Blue contour signifies $\lambda$-ensemble, while yellow corresponds to $N$ or $G_{\rm N}$-ensemble in field theory and supergravity. Green arrows signify a change of ensemble.}
	\label{fig:1}
\end{figure}
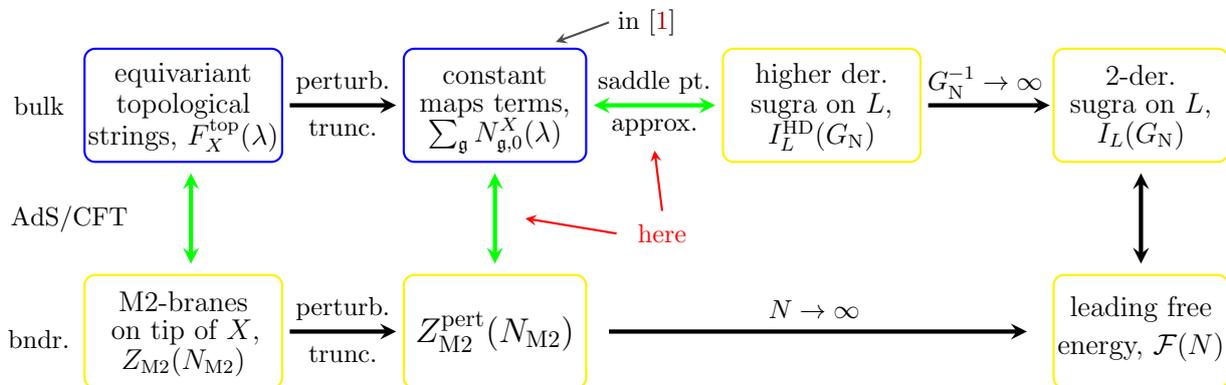

In \cite{Cassia:2025aus}, we put forward the following broad conjecture. We proposed that equivariant topological string theory, defined on a non-compact toric manifold $X$, provides the exact gravitational description of D- and M-branes with a supersymmetric transverse space $L$. In the simplest case of M2-branes, which we primarily focus on here, $X$ should be understood as the resolution of the cone $C(L)$, see also \cite{Martelli:2023oqk,Colombo:2023fhu}. The supersymmetry condition requires that $L$ be a Sasakian space, which in turn ensures that $X$ satisfies the Calabi--Yau condition. For a more general introduction to the subject of equivariant topological strings and the context of the above conjecture, we refer the reader to \cite{Cassia:2025aus}. In that work, we considered an equivariant extension of the perturbative topological string free energy as a function of the so-called redundant K\"ahler parameters, $\lambda$, and equivariant parameters, $\e$:~\footnote{More generally, in \cite{Cassia:2025aus} we conjectured that a fully non-perturbative extension of the present proposal should exist, see the left-most part of Fig.\ \ref{fig:1}. At present, however, the requisite framework for defining equivariant topological strings in sufficient generality is not yet available, and consequently no direct evidence can be provided for such a claim. In this work we therefore adopt a pragmatic standpoint and restrict our analysis to the perturbative sector, where the proposal can be formulated and tested with high precision. We refer the reader to \cite{Cassia:2025aus} for further discussion on the current status and open challenges of equivariant topological string theory.}
\be
\label{eq:pertstring}
    F^\text{top,pert}_X (\lambda, \epsilon; g_s) = \sum_{\mathfrak{g} = 0}^\infty g_s^{2 (\mathfrak{g}- 1)}\,  N^X_{\mathfrak{g}, 0} (\lambda, \e)\ ,
\ee
with $g_s$ the string coupling constant and $\mathfrak{g}$ the genus of the corresponding contribution.

In the present work, we elaborate on and verify the holographic relation described above by expanding upon the schematic formula proposed in \cite{Cassia:2025aus}, see Fig.\ \ref{fig:1},~\footnote{Note that very recently, in \cite{Gautason:2025plx}, appeared an interesting proposal to define supergravity directly in the $\lam$-ensemble, which suggests a further expansion of the schematic diagram here.}
\be
\label{eq:mainconjecture-pert}
	Z^\text{pert}_L (\e, N_{\rm M2}) = \int {\mathd} \lambda\, \exp \left(  F^\text{top,pert}_X (\lambda, \epsilon; g_s)- \lambda\, N_{\rm M2}  \right)\ ,
\ee
where $N_{\rm M2}$ is the exact M2-brane charge that differs from the number of branes and dual gauge group rank $N$ by a constant shift, see below. This proposed relation, spelled out explicitly in \eqref{eq:mainconjecture-pert-round} and \eqref{eq:mainconjecture-pert-squashed}, allows us to test holography at \emph{finite} $N$ for all M2-brane models that relate to toric manifolds, including ABJM theory, \cite{Aharony:2008ug}, and a large set of more general quiver models, see \cite{Benini:2009qs,Cremonesi:2010ae} for the construction and relation between quiver and toric diagrams.

Using the full knowledge of the equivariant constant maps $N^X_{\mathfrak{g}, 0}$ from \cite{Cassia:2025aus}, which can be computed from the equivariant volume of the manifold,~\footnote{Here $\omega_\lam-\e_iH^i$ is an equivariant completion of the K\"ahler form of $X$ parametrized by redundant variables $\lam^i$.}
\be
 \BV_X (\lam, \e) = \int_X\mathe^{\omega_\lam-\e_iH^i}\,,
\ee
we demonstrate that the right-hand side of \eqref{eq:mainconjecture-pert} precisely evaluates to the Airy function formulation of the holographically dual (possibly squashed) three-sphere partition function of the corresponding M2-brane model, see \cite{Kapustin:2009kz,Drukker:2010nc,Herzog:2010hf,Santamaria:2010dm,Jafferis:2010un,Hama:2010av,Hama:2011ea,Martelli:2011qj,Cheon:2011vi,Jafferis:2011zi,Fuji:2011km,Imamura:2011wg,Marino:2011eh,Martelli:2011fu,Hatsuda:2012dt,Hatsuda:2013oxa,Moriyama:2014gxa,Nosaka:2015iiw,Hatsuda:2016uqa,Kubo:2019ejc,Hosseini:2019and,Kubo:2020qed,Chester:2021gdw,Bobev:2022jte,Bobev:2022eus,Geukens:2024zmt,Kubo:2024qhq,Kubo:2025jxi,Bobev:2025ltz,Kubo:2025dot} and references therein. At the same time, we also provide a detailed analysis of the relation between the equivariant topological string constant map terms and the effective 4d higher-derivative (HD) supergravity description, which has been considered in \cite{Bobev:2020egg,Bobev:2021oku,Hristov:2021qsw,Hristov:2022lcw,Hristov:2022plc,Hristov:2024cgj}. In addition to the diagram above, explaining the role of the $\lambda$ parameters, we must also clarify the role of the equivariant parameters $\epsilon$ in their connection to supergravity and field theory. These relations can be summarized as follows:
\be
\label{eq:hologravgeomatch}
\begin{array}{ccccc}
	\epsilon_i & \leftrightarrow & X^I & \leftrightarrow & \Delta_i \\
    (\text{geometry}) && (\text{sugra}) && (\text{SCFT}) \\
\end{array}
\ee
where $X^I$ are the complex scalars in (off-shell) supergravity, \cite{Andrianopoli:1996cm,Lauria:2020rhc}, while the $\Delta_i$ are supersymmetric deformations of the dual field theory, \cite{Jafferis:2010un,Jafferis:2011zi}. 

To ensure that the above Laplace transform in \eqref{eq:mainconjecture-pert} is well-defined and to derive the claimed result, we proceed in several steps. First, we verify the leading order expressions on both sides of the holographic correspondence in Section~\ref{sec:2}. This is done by analyzing the genus-zero constant maps term and the large $N$ behavior of the squashed sphere partition function, in agreement with \cite{Martelli:2023oqk} and previous related references, \cite{Couzens:2018wnk,Gauntlett:2018dpc,Hosseini:2019use,Hosseini:2019ddy,Gauntlett:2019roi,Kim:2019umc,Boido:2022mbe}. A key input in this step is the mesonic twist of the $\lam$ parameters (i.e.\ the blow-down of all non-trivial two-cycles in $X$), which is implemented at the level of the equivariant volume of $X$. This twist eliminates the baryonic directions and consequently simplifies the equivariant volume to the following expression, as detailed in \cite{Cassia:2025aus}:
\be
\label{eq:mesonicequivol}
 \BV_X (\lam_\text{mes.}, \e)
% = \mathe^{\e_i \lam^i_\text{mes.}}\, \VVV^\text{mes.}_X (\e)
 = \mathe^{\e_i \lam^i_\text{mes.}}\, C_X (\e)\ ,
\ee
where we introduced the more familiar quantity $C_X(\e):=\BV_X(0,\e)=\VVV^\text{mes.}_X(\e)$, which is a homogeneous function of degree $-\dim_\BC(X)$ in the $\e$ variables, and is commonly used on the field theory side of the calculation in the conventions of \cite{Marino:2011eh,Nosaka:2015iiw,Hatsuda:2016uqa}.~\footnote{Additionally, note that the variables $\e$, although natural to use holographically, are redundant and can be replaced with the independent parameters $\nu$ as in \cite{Cassia:2025aus}. This redundancy has an exact analog in the dual supergravity variables $X$ and field theory variables $\Delta$, in accordance with \eqref{eq:hologravgeomatch}.} Using the mesonic twist condition, we are both able to establish the relations \eqref{eq:hologravgeomatch} and justify the universal parametrization, (see \cite{Cassia:2025aus})
\be
	\lambda_\text{mes.}^i = \tilde \mu\ , \quad \forall i\ ,
\ee  
which allows us to reduce the contour integral \eqref{eq:mainconjecture-pert} to a single complex integration variable.

Next, in Section~\ref{sec:3}, we turn to the subleading corrections arising from the genus-one constant maps terms. At this stage, we must consider the possible refinement of the topological string partition function, which is feasible on toric manifolds, through the refined topological vertex, see \cite{Aganagic:2003db,Okounkov:2003sp,Iqbal:2007ii, Huang:2010kf, Krefl:2010fm}, using the Nekrasov partition function, \cite{Nekrasov:2002qd,Nekrasov:2003rj}. The refinement, parametrized by $\mathfrak{b}$, introduces an additional term at the genus-one constant maps level (along with a large number of terms at higher genus). We can express the refinement at genus-one level as the deformation
\be
	N^X_{1, 0} (\lambda_\text{mes.}, \epsilon) \mapsto  N^X_{1, 0} (\lambda_\text{mes.}, \epsilon) - (\mathfrak{b} + \mathfrak{b}^{-1})^2\, M^X_{1, 1, 0} (\lambda_\text{mes.}, \epsilon)\ ,
\ee
where, in our conventions, the unrefined limit corresponds to $\mathfrak{b}=\sqrt{-1}$. The deformation term $M^X_{1, 1, 0}$ has no obvious geometric origin. 
Nevertheless, we are able to define it in terms of the triple intersection numbers of the manifold $X$, incorporating the additional input from the higher-derivative supergravity building blocks, viewed from a 4d perspective. To achieve this, we first dedicate Section~\ref{sec:intermezzo} to a selective summary of supergravity results from \cite{Hristov:2021qsw,Hristov:2024cgj}, before presenting the refined genus-one contribution and its holographic match to subleading $N$ corrections. In the process we establish the relation with HD supergravity depicted on Fig.\ \ref{fig:1}, which requires us to identify the refinement parameter, $\mathfrak{b}$, with the squashing parameter of the three-sphere, $\squb$.~\footnote{It is important to emphasize that, in the present conventions, the unrefined limit corresponds to $\mathfrak{b} = \mathi$. After the identification $\mathfrak{b} = \squb$,  this does \emph{not} coincide with the limit of vanishing squashing (round sphere), $\squb = 1$. It is therefore crucial to include refinement already for the round sphere partition function.}

Finally, in Section~\ref{sec:5} we turn to the exact evaluation of the integral in \eqref{eq:mainconjecture-pert}, ignoring the constant contributions from the higher-genus constant maps terms that can be factorized outside. We give more details for different M2-brane models and their match with localization results in the respective section, but can already summarize the main result in the following compact formula that holds up to a constant $\cO (N^0)$ prefactor and non-perturbative corrections:
\be
\label{eq:centralresult}
	Z_{S^3_\squb}^\text{pert} (\Delta, \squb, N_{\rm M2}) \simeq {\rm Ai} \Big[ \left( \frac{2\, C_{X} (\Delta)}{\pi^2\, (\squb+\squb^{-1})^4}\right)^{-1/3} \left(N_{\rm M2}- \frac{ C_{X} (\Delta)}{12}\, \left(\frac{2\, k_2 (\Delta)}{(\squb+\squb^{-1})^2} - k_3 (\Delta) \right) \right) \Big]\ ,
\ee
where the $\Delta_i$'s parametrize the mesonic deformations on the three-sphere, obeying the supersymmetric constraint
\be
	\sum_i \Delta_i = 2\ .
\ee
In the previous formula we used the following short-hands
\be
\label{eq:shorthandc2andc3}
	k_2 (\e) := \frac1{C_X (\e)}\, \sum_{i < j} \frac{\partial^2 \BV_X (\lam,  \e)}{\partial \lam^i \partial \lam^j} \Big|_{\lam = 0}\ , \qquad k_3 (\e) :=\frac1{C_X (\e)} \sum_{i < j < k} \frac{\partial^3 \BV_X (\lam,  \e)}{\partial \lam^i \partial \lam^j \partial \lam^k} \Big|_{\lam = 0}\ ,
\ee
for two functions which come from the equivariant integrals of the second and third Chern classes of $X$, respectively. Likewise, the exact M2-brane charge is related to the equivariant volume via the Euler characteristic,~\footnote{Here we assume for simplicity no discrete torsion contribution, see \cite[Appendix~A]{Cassia:2025aus} for details. Note that the shift in the exact M2-brane charge does \emph{not} come from equivariant integration, but we can again express it in terms of equivariant intersection numbers.}
\be
\label{eq:exactchargeviaEulerchar}
	N_{\rm M2} = N - \frac{\chi (X)}{24} = N - \frac1{24} \sum_{i < j < k < l} \frac{\partial^4 \BV_X (\lam,  \e)}{\partial \lam^i \partial \lam^j \partial \lam^k \partial \lam^l} \Big|_{\lam = 0}\ ,
\ee
using that $X$ is a four-fold. Notice that \emph{all} quantities in \eqref{eq:centralresult} can be derived from the equivariant volume $\BV_X (\lam, \e)$, which encodes the information about the classical geometry.

The integral representation and asymptotic expansion of the Airy function reads
\be
\label{eq:airyasympt}
\text{Ai} (z) = \frac1{2 \pi\mathi} \int_\cC \exp \left(\frac{t^3}{3} - z t \right)\, \mathd t \sim \frac{ {\color{purple} \mathe^{-2/3\, z^{3/2}} } }{2 \sqrt{\pi} z^{1/4}}\,
 \sum_{n=0}^\infty \frac{(-1)^n 3^n \Gamma(n+\frac56) \Gamma(n+\frac16)}{2 \pi n!\, 4^n\, z^{3n/2} }  \ , 
\ee
where the contour $\cC$ starts at infinity with argument $-\pi/3$ and ends at infinity with argument $\pi/3$. Note that we have colored the exponent above to illustrate the relation between the topological string and the supergravity frameworks. We further define the perturbative free energy of a given M2-brane theory as the saddle point evaluation of \eqref{eq:mainconjecture-pert},
\be
\label{eq:deffreenergy}
	\cF_{S^3_\squb} (z) := - \left.\log Z_{S^3_\squb}^\text{pert} (z)\right|_{\text{saddle pt.}}\ ,
\ee
where we identify the variable $z$ as the argument of the Airy function in \eqref{eq:centralresult}. We are then going to show that the HD supergravity action, evaluated on the asymptotically AdS$_4$ background with squashed $S^3$ boundary, $\cI_{S^3_\squb}$, corresponds to,~\footnote{Note that this equation does not mean the supergravity framework can only give the leading exponential behavior of the dual partition function. The quantity $\cI_{S^3_\squb}$  is the strictly classical higher-derivative supergravity action, not incorporating 1-loop and non-perturbative effects, which could relate to the $\log$ and further corrections here, see \cite{Bhattacharyya:2012ye,Liu:2017vbl,Hristov:2021zai,Bobev:2023dwx,Gautason:2023igo,Beccaria:2023ujc,Beccaria:2023sph}.}
\be
\label{eq:HDsugractiongeneral}
	\cI_{S^3_\squb} (z) = {\color{purple} \frac23\, z^{3/2} }\ , \quad \Rightarrow \quad \cI_{S^3_\squb} (z) =  \cF_{S^3_\squb} (z)  = - \log Z_{S^3_\squb}^\text{pert} (z) - \frac14\, \log z + ... \ .
\ee

Equation \eqref{eq:centralresult} constitutes the central result of the present paper, which, for the case of M2-branes on the tip of $\BC^4/\BZ_k$, or ABJM theory, was conjectured in \cite{Hristov:2022lcw,Bobev:2022jte} based on the superposition of constraints from field theory \cite{Fuji:2011km,Marino:2011eh,Nosaka:2015iiw,Hatsuda:2016uqa,Chester:2021gdw} and supergravity \cite{Hristov:2021qsw}. We should note that the last term proportional to $k_3(\Delta)$, which comes from the refinement, has been partially fitted (passing all the consistency checks within our framework) based on this agreement. On the other hand, the general formula is supposed to apply to all M2-brane models on toric manifolds, which extends vastly beyond ABJM theory itself to a large number of different Chern--Simons-matter quivers, \cite{Benini:2009qs,Cremonesi:2010ae}.

In addition to the squashed sphere result, in Section~\ref{sec:6}, we explore the extension of the conjectured relation \eqref{eq:mainconjecture-pert} to the case of M2-branes wrapped on toric surfaces $\Sigma$, such as spindles and two-spheres. The geometric picture, proposed again in \cite{Martelli:2023oqk} and further developed in \cite{Cassia:2025aus}, suggests that we must consider the fibration of the space $X$ over $\Sigma$. Holographically, this construction corresponds to the respective twisted or anti-twisted spindle index, generalizing the twisted and superconformal indices on the two-sphere, respectively, see \cite{Kim:2009wb,Kapustin:2011jm,Beem:2012mb,Benini:2015noa,Benini:2015eyy,Hosseini:2022vho,Bobev:2023lkx,Bobev:2024mqw, Inglese:2023wky, Colombo:2024mts} and references therein. Unlike the three-sphere case, we were only able to affirmatively test the relation at leading order, reproducing the results of \cite{Martelli:2023oqk} and previous references, \cite{Couzens:2018wnk,Gauntlett:2018dpc,Hosseini:2019use,Hosseini:2019ddy,Gauntlett:2019roi,Kim:2019umc,Boido:2022mbe}. We also find a qualitative agreement with the conjecture of \cite{Hristov:2022lcw}, which suggests that the answer factorizes in terms of two different Airy functions. However, we did not explore the refinement of the genus-one constant maps in this case, leaving the exact \emph{finite} $N$ prediction for future studies. Our results thus summarize the current state of the art regarding the wrapped M2-brane construction while leaving intriguing open questions for future investigation.

Lastly, in Section~\ref{sec:7} we look at D3-brane models that can be possibly described by an analogous conjectural formula as \eqref{eq:mainconjecture-pert}. We suggest that they are similarly described by a Laplace transform over the $\lam$ parameters, which this time involves the double intersection numbers of the underlying manifold. We briefly apply our reasoning at leading order, in agreement with \cite{Couzens:2018wnk,Gauntlett:2018dpc,Hosseini:2019use,Hosseini:2019ddy,Gauntlett:2019roi,Boido:2022mbe,Martelli:2023oqk}. We again leave the precise relation for future study, and finish with a broader set of open questions in Section~\ref{sec:conclusion}.

\section{Leading order in \texorpdfstring{$N$}{N}: flux vs. \texorpdfstring{$\lam$}{lambda} ensembles}
\label{sec:2}
We are going to consider here the leading order contributions to the M2-brane background in 11d supergravity and the related equivariant topological string construction, together with the holographically dual results in the large $N$ approximation. One can think of this section as a continuation of Sec.~6 in \cite{Cassia:2025aus}. In particular, we focus on the near-horizon of spacetime filling M2-branes, corresponding holographically to Chern--Simons-matter theories on the round $S^3$ background and its squashed generalization. In order to simplify the reading experience and not refer constantly to the companion paper, we present a short summary below, which allows us to build further on the relation between the two ensembles of interest. Our main goal here is to revisit the known large $N$ behavior of the dual partition function and relate it to the genus-zero constant maps term, allowing us to set up explicitly the proposed integration contour in \eqref{eq:mainconjecture-pert}.

\subsection*{Summary of M2-brane background}
M2-branes exhibit an AdS$_4$ near-horizon geometry with an internal Sasakian space $L$,
\be
	M_{11} = \mathrm{AdS}_4 \times L\ ,
\ee
where the radii of the two factors are proportional to each other and are given by a positive power of the M2-brane charge, given at leading order by
\be
	N = \frac1{(2 \pi l_P)^6} \int_L \star F_4\ ,
\ee 
where $F_4$ is the field strength of the 11d supegravity 3-form gauge field (c.f.\ \cite[Section~6]{Cassia:2025aus}). The dual holographic description of this system involves M2-branes on the tip of the cone $X = C (L)$, which is in general a 3d $\cN=2$ CS-matter quiver theory with a gauge group of rank $N$. Most importantly, as discussed in \cite{Couzens:2018wnk,Gauntlett:2018dpc,Hosseini:2019use,Hosseini:2019ddy,Gauntlett:2019roi}, the  R-charge assignments $\Delta_i$ can be related to the leading topological string expression via~\footnote{Note that we do not strictly speaking need to stay at the mesonic locus $\lam=\lam_{\rm mes.}$. The equation would still hold true, but would include more general baryonic symmetries, see \cite{Hosseini:2025mgf}.}
\be
\label{eq:holoDelta}
	\frac1{4 \pi^2}\,  \left. \frac{\partial N^X_{0, 0} (\lambda, \epsilon)}{\partial \lambda^i} \right|_{\lam=\lam_{\rm mes.}} = \Delta_i\, N \ , \quad \forall i\ ,
\ee
where the genus-zero constant maps are defined as
\be
\label{eq:holocond1}
	N^X_{0,0} (\lam, \e) = \frac16\, \sum_{i, j, k} \frac{\partial^3 \BV_X (\lam,  \e)}{\partial \lam^i \partial \lam^j \partial \lam^k} \Big|_{\lam = 0}\, \lam^i \lam^j \lam^k\ ,
\ee
which also corresponds to the degree 3 term in the $\lam$-expansion of $\BV_X (\lam,  \e)$.
Supersymmetry of the dual field theory imposes the condition
\be
\label{eq:holocond2}
	\sum_i \Delta_i = 2\ .
\ee
We turn to discuss the meaning of $\lam_{\rm mes.}$ next. Note that \eqref{eq:holoDelta} applies only in the case of pure AdS$_4$ background in $M_{11}$ without geometric deformations, corresponding holographically to field theory on the round $S^3$, where the $\Delta$ parameters play the role of (complexified) masses for the matter multiplets.

\subsection*{Mesonic twist}
\label{sec:2.2}
As already noted in \cite[Section~2.3]{Cassia:2025aus} based on \cite{Hosseini:2019use,Hosseini:2019ddy}, it makes sense holographically to impose the so called mesonic twist on the toric geometry $X$, effectively shrinking all non-trivial two-cycles in homology and keeping only the toric action related to the global isometry of the manifold. In short, this entails setting all ``effective'' Kähler parameters to zero, which at the level of the $\lam$ parameters corresponds to the condition that
\be
\label{eq:mesonictwist}
	\sum_i \lambda_\text{mes.}^i\, Q_i^a = 0\ , \quad \Rightarrow \quad \lambda_\text{mes.}^i = \sum_\alpha \mu^\alpha\, v^i_\alpha\ ,
\ee
where $v_\alpha$ are the vectors in the toric fan, which span the co-kernel of the charge matrix $Q$, i.e.\ $\sum_iQ_i^a v_\alpha^i=0$. We can effectively think of the new variables $\mu$ as the ``reduced'' $\lam$ variables, i.e.\ those ineffective Kähler parameters which remain linearly independent after setting all effective Kähler parameters to zero. We refer to this procedure as the \emph{mesonic twist}.

This allows us to write down the general behavior,
\be
\label{eq:mesonictwistvolume}
 \BV_X (\lam_\text{mes.}, \e) = \mathe^{\e_i \lam^i_\text{mes.}}\, \VVV^\text{mes.}_X (\e)\ ,
% \qquad C_X (\e) := \VVV_X^\text{mes.} (\e)\ ,
\ee
as already anticipated in \eqref{eq:mesonicequivol}, where $\VVV_X(t,\e)=\int_X\mathe^{\omega_t-\e_iH^i}$ is the version of the equivariant volume parametrized by effective K\"ahler variables $t^a$, which vanish at the mesonic twist locus, hence $\VVV_X^\text{mes.}(\e):=\VVV_X(0,\e)=C_X(\e)$.
Using the same mesonic twist condition on the $\lam$ parameters, we further derived in \cite{Cassia:2025aus} a simplified expression for the first two constant maps terms,
\be
\label{eq:mesonicconstmaps}
\begin{aligned}
 N_{0,0}^\text{mes.} (\lam_\text{mes.}, \e) &=
 \frac16\, \left(\e_i \lam_\text{mes.}^i\right)^3\, C (\e)\ , \\
 N_{1,0}^\text{mes.} (\lam_\text{mes.}, \e) &=
 \frac1{24}\,
 \left(\e_i \lam_\text{mes.}^i\right)\, \sum_{i < j} \frac{\partial^2 \BV (\lam,  \e)}{\partial \lam^i \partial \lam^j} \Big|_{\lam = 0}\,\ ,
\end{aligned}
\ee
where we suppressed the index $X$, which is implied.~\footnote{ Note a possible confusion regarding the term \emph{mesonic twist} could arise in comparison with \cite{Hosseini:2019ddy}. We only use this term to signify the condition on $\lam$, \emph{not} the condition on $\e$ discussed next.}

In addition to the mesonic twist, one can define the so called \emph{mesonic constraint} on the equivariant parameters $\e$, which amounts to the requirement that the mesonic twist commutes with the $\lam$-derivatives, see again \cite[Section~2.3]{Cassia:2025aus}. In practice, we have the equations
\be
\label{eq:mesonicconstraint}
	 \e_i \stackrel{!}{=}
 \left.\frac{\partial}{\partial\lam^i}\log\BV(\lam,\e)\right|_{\lam=\lam_{\rm mes.}}\ ,
\ee
which is to be understood as a set of non-linear constraints for the $\e$ variables.
Upon solving this system of equations, we are able to fix some but not all of the $\e$ parameters in terms of the remaining ones, which can be regarded now as ``effective'' equivariant parameters. This procedure allows then to lift the redundancy in the $\e$.

We are going to discuss below how the mesonic constraint has a dual analog in the language of F-extremization, \cite{Jafferis:2010un,Jafferis:2011zi}. Note that this constraint on $\e$ is a particular choice that one can make, but it has so far no fundamental geometrical significance. As explained in \cite{Cassia:2025aus}, as well as in \cite{Martelli:2023oqk,Colombo:2023fhu}, we could replace the parameters $\e_i$ with the ``effective'' equivariant parameters $\nu_\alpha$, satisfying $\nu_\alpha = \sum_i v^i_\alpha \e_i$. We do not use these variables here for the sociological reason that available dual field theory calculations in literature can be more directly compared to the present results using the $\e_i$, being immediately related to the $\Delta_i$ as discussed in due course, see again \cite{Jafferis:2011zi}.

\subsection{Round \texorpdfstring{$S^3$}{S3} holography}
The simplest background that can be obtained as the near-horizon limit of multiple M2-branes (see again the summary in \cite[Section~6]{Cassia:2025aus}) is the AdS$_4 \times L$ background with the four-form field strength proportional to the volume form on AdS$_4$. In this case, the metric preserves the maximal possible isometry on both spaces, in particular at least $\cN=2$ supersymmetry from the four-dimensional perspective (corresponding to 8 real supercharges). The best understood holographic interpretation of this solution, which necessitates the Wick rotation to Euclidean signature, is when we choose the boundary to have a three-sphere topology.~\footnote{One can choose other boundary slicings that preserve supersymmetry such as $S^1 \times \Sigma$, but then empty AdS$_4$ becomes a subleading saddle. We discuss such a case involving black hole backgrounds in Section~\ref{sec:6}.} The maximal isometry of (Euclidean) AdS$_4$ (or simply $\mathbb{H}_4$) then dictates that the induced metric on the sphere is the round one.

Combining conditions \eqref{eq:holoDelta} and \eqref{eq:holocond2}, holography dictates that the R-charge assignments $\Delta_i$ are given geometrically by
\be
\label{eq:deltai-holo}
	\Delta_i = 2\,  \frac{\left( \partial N^X_{0, 0} (\lambda, \epsilon) / \partial \lam^i\right) |_{\lam = \lam_\text{mes.}}}{\left( \sum_j \partial N^X_{0, 0} (\lambda, \epsilon) / \partial \lam^j\right) |_{\lam = \lam_\text{mes.}}}\ .
\ee
By the definition of the equivariant volume as generating function of equivariant intersection numbers, we find
\be
 \left.\frac{\partial N^X_{0,0}(\lam,\e)}{\partial\lam^i}\right|_{\lam=\lam_\text{mes.}}
 = \frac{(\e_j\lam^j_\text{mes.})^2}{2}
 \left(\left.\frac{\partial\BV(\lam,\e)}{\partial\lam^i}\right|_{\lam=0}\right)\ .
\ee
Upon imposing the mesonic constraint~\footnote{Had we included the baryonic directions, as in \cite{Hosseini:2025mgf}, we would only be able to invert the above relation by a general function $\Delta_i (\lam, \e)$, which would in turn make it impossible to give as explicit final results as the ones that follow. Nevertheless, it is conceptually straightforward to repeat the following steps without the mesonic twist, resulting in a more complicated integration contour. We leave this generalization for the future, as it is not currently clear whether the dual field theory feels the baryonic symmetries.}, \eqref{eq:mesonicconstraint}, we can further derive
\be
 \left.\frac{\partial N^X_{0,0}(\lam,\e)}{\partial\lam^i}
 \right|_{\substack{\lam=\lam_\text{mes.}\\+\eqref{eq:mesonicconstraint}}}
 = \frac{(\e_j\lam^j_\text{mes.})^2}{2}\,
 \e_i \, C_X(\e)
\ee
which, after plugging back into \eqref{eq:deltai-holo}, gives
\be
\label{eq:mesonictwistimplication}
	\Delta_i = 2\, \frac{\e_i}{\sum_j \e_j}\ ,
\ee
so that $\Delta_i$ and $\e_i$ must be proportional to each other. Assuming that their sums are normalized in the same way, i.e.\ $\sum_i\e_i=2=\sum_i\Delta_i$, we deduce that
\be
 \Delta_i = \e_i\ .
\ee
The direct relation between the parameters $\Delta_i$ and $\epsilon_i$ allows us to compare the field theory localization results against the topological string predictions after taking into account the change of ensemble between $\lambda$ and $N_{\rm M2}$ variables.~\footnote{Although we explicitly used the mesonic condition \eqref{eq:mesonicconstraint} in order to relate $\e_i$ to $\Delta_i$ holographically, this is due to the specific field theory parametrization dictating \eqref{eq:holocond1}. One is however free to reparametrize the R-charges $\Delta$ in a arbitrary way by homogeneous functions of first degree. This freedom is analogously followed by the $\e_i$ parameters geometrically if we relax the condition \eqref{eq:mesonicconstraint}. Therefore we believe that the identification between $\Delta$ and $\e$ can be assumed to hold in full generality, irrespective of particular choice for fixing the redundancy in the parametrization of the $\e$ parameters.}

Next, we notice that there is one last holographic relation at leading order, 
\be
\label{eq:finalholocond1}
\frac1{4 \pi^2}\, \sum_i \left. \frac{\partial N^{X}_{0, 0} (\lam, \e) }{\partial \lambda^i} \right|_{\lam=\lam_{\rm mes.}} = 2\, N\ ,
\ee
which is not captured by the identification between $\epsilon_i$ and $\Delta_i$, \eqref{eq:mesonictwistimplication}. Thus the original $\lam$ extremization of the genus-zero constant maps, \eqref{eq:holocond1}, can now be considered as a single equation. In accordance with the discussion in \cite[Section~2.3]{Cassia:2025aus}, we can define a universal choice of parametrization for the solutions of \eqref{eq:mesonictwist} in terms of the variables $\mu^\alpha$. This universal choice corresponds to fixing
\be  
\label{eq:univparametrization}
\mu^1 = \tilde \mu\ , \quad \mu^{\alpha > 1} = 0\ ,  \quad \Rightarrow \quad \lambda_\text{mes.}^i = \tilde \mu\ , \quad \forall i\ , 
\ee  
where we used that the first vector in the toric fan is conventionally chosen as $v_1=(1,1,\dots,1)$.
%This provides a choice of universal parametrization, in the sense that it always satisfies the mesonic twist condition.
Using this universal parametrization in terms of $\tilde\mu$, we can rewrite \eqref{eq:finalholocond1} in the following way:
\be
\label{eq:extremization-mu-tilde}
	\frac1{4 \pi^2}\, \frac{\partial N^{X, \text{mes.}}_{0, 0} (\tilde \mu, \e)}{\partial \tilde \mu} = 2\, N\ ,
\ee
where $N^{X,\text{mes.}}_{0,0}(\mu,\e)=N^X_{0,0}(\lam,\e)|_{\lam^i=\mu^\alpha v^i_\alpha}$.

Furthermore, we can explicitly write the mesonic genus-zero constant maps term, c.f.\ \eqref{eq:mesonicconstmaps},~\footnote{We note that the universal parametrization we have chosen, \eqref{eq:univparametrization}, is perfectly aligned with the holographic constraint on the $\Delta$, and respectively $\e$, parameters, \eqref{eq:holocond2}. Given the previous footnote, which suggests that the constraint \eqref{eq:holocond2} is a conventional choice, we expect that the $\lam_\text{mes.}$ parametrization needs to always be coordinated with the supersymmetry constraint on the $\Delta_i$.}
\be
\label{eq:mesonicconstmapsroundsphereinmu}
	N^{X, \text{mes.}}_{0, 0}  (\tilde \mu, \e) =  \frac43\, C_X (\e)\, \tilde \mu^3\ .
\ee
Let us now recall the holographic formula for the large $N$ expression of the free energy, which can be checked explicitly on a number of M2-brane models on a Sasakian manifold $L$, see \cite{Drukker:2010nc,Herzog:2010hf,Santamaria:2010dm,Martelli:2011qj,Cheon:2011vi,Jafferis:2011zi},
namely
\be
	\cF_L (\Delta, N) = \frac{4 \sqrt{2} \pi}3\, \frac{N^{3/2}}{\sqrt{C_X (\Delta)}}\ .
\ee
We observe that we can derive the leading order M2-brane partition function in the saddle-point approximation of the following integral,
\be
	Z_L^\text{pert} (\Delta, N) = \int \mathd \tilde \mu\, \exp \left(\frac1{4 \pi^2}\, N^{X, \text{mes.}}_{0, 0}  (\tilde \mu, \Delta) -  2\, N\, \tilde \mu \right)\ .
\ee
The saddle-point equation gives precisely \eqref{eq:extremization-mu-tilde}, and solving w.r.t.\ $\tilde\mu$ we obtain
\be
	- \log Z_L^\text{pert} (\Delta, N) \stackrel{\text{saddle pt.}}{\approx}  \frac{4 \sqrt{2} \pi}3\, \frac{N^{3/2}}{\sqrt{C_X (\Delta)}}\ .
\ee
This is in agreement with the leading behavior of the free energy of M2-brane theories on the round three-sphere above, and serves to precisely fix the normalization factors between the sole remaining parameter $\tilde \mu$ and its conjugate flux $N$.
%\be
%	\tilde \mu \leftrightarrow N\ .
%\ee

\subsubsection*{Conclusion}
Based on the above considerations, we are ready to specify formula \eqref{eq:mainconjecture-pert} for the round sphere case more concretely:
\be
\label{eq:mainconjecture-pert-round}
	Z^\text{pert}_L (\Delta, N_{\rm M2}) = \int \mathd \tilde \mu\, \exp \left(  F^\text{top,pert}_X (\lambda^i = v^i_1\, \tilde \mu, \epsilon_i = \Delta_i; g_s = 2 \pi)- 2 N_{\rm M2}\, \tilde \mu  \right)\ ,
\ee
where we used that the exact M2-brane charge $N_{\rm M2}$ in the large $N$ limit is simply identified with $N$ itself, see \cite[Appendix~A]{Cassia:2025aus} for more details. Anticipating our results, we are going to assume that the integration contour for the $\tilde \mu$ variable coincides with the one needed to arrive at the integral representation of the Airy function, \eqref{eq:airyasympt}.

\subsection{Squashed \texorpdfstring{$S^3_\squb$}{S3b} generalization}
In the spirit of allowing arbitrary supersymmetric deformations which preserve only the topology of the background, we can consider a more general metric both on the internal space $L$ and the external space $\mathbb{H}_4$. As discussed in \cite{Cassia:2025aus}, the supersymmetric deformations of $L$ are those that preserve the Sasakian condition. This corresponds to a constraint on the Reeb vector $\xi$, given by
 \be
	\xi =2 \pi\, \sum_i \epsilon_i\, \partial_{\varphi_i}\ ,
\ee
where the isometries $\partial_{\varphi_i}$ correspond to the toric action on the cone $X$ over $L$. The equivariant parameters $\e_i$ then correspond to the internal deformations preserving the topology and a fraction of the supersymmetries.
In the 4d supergravity description, the internal deformations manifest themselves in non-trivial profiles for the scalars $X^I$, while in field theory they appear as $\Delta_i$ deformations. 

Additionally, in the case of $S^3$ boundary for the external space, we are allowed to introduce one more deformation that preserves supersymmetry, and it corresponds to squashing the round metric on $S^3$ such that one breaks the $\SO(4)$ isometry group to the maximal torus $U(1) \times U(1)$, \cite{Plebanski:1976gy,Alonso-Alberca:2000zeh,Martelli:2011fu}. This deformation in itself also breaks the conformal symmetry, and can be seen geometrically in the supergravity solution. Even though the squashing of the external space cannot be seen geometrically on the internal manifold $L$ and its cone $X$, we find that the squashing parameter enters the constraint between the equivariant parameters, via supersymmetry. The underlying background was called $\Omega \mathbb{H}_4$ in \cite{Hristov:2022plc}, and we discuss it in Section~\ref{sec:OmegaBack}.

In this case, we find that supersymmetry imposes the constraint
\be
\label{eq:susy-constr-b}
 \sum_i \epsilon_i = (\squb + \squb^{-1})\ ,
\ee
where $\squb$ is the squashing parameter\footnote{The conical deficit parameter $b$ of the spindle $\WPL$ is not to be confused with the squashing parameter $\squb$ on the squashed sphere, here and elsewhere in this paper.} of the three-sphere at the boundary of $\mathbb{H}_4$. The above constraint generalizes to arbitrary squashing the relation \eqref{eq:mesonictwistimplication}, which holds at $\squb=1$. Again, we note that the constraint among $\e_i$ does not have a clear geometric origin for us, but can be derived either via holography or using the supergravity background, see Section~\ref{sec:OmegaBack}.

Given the ``deformed'' supersymmetry constraint \eqref{eq:susy-constr-b}, we can repeat the analysis of the previous section to find
\be
 N^{X, \text{mes.}}_{0, 0} (\tilde \mu, \e)
 = \frac{\tilde \mu^3}{6}\,(\squb+\squb^{-1})^3\, C_X (\e)\ .
\ee
The R-charge constraint in the presence of squashing is unchanged, \cite[Eq.~(2.17)]{Hosseini:2019and},
\be
\label{eq:Deltaconstr}
	\sum_i \Delta_i = 2\ ,
\ee
which leads to the deformed identification of parameters
\be
\label{eq:squashedepsilontob}
    \e_i = \frac{(\squb + \squb^{-1})}2\, \Delta_i\ .
\ee
Therefore, we have
\be
 N^{X, \text{mes.}}_{0, 0} (\tilde \mu, \e)
 = \frac{\tilde \mu^3}{6}\,(\squb+\squb^{-1})^3\, C_X ((\squb+\squb^{-1}) \Delta/2)
 = \frac{8\, \tilde \mu^3}{3\, (\squb+\squb^{-1})}\, C_X (\Delta)\ ,
\ee
which naturally generalizes \eqref{eq:mesonicconstmapsroundsphereinmu}. In the latter equality, we used the property that $C_X(\e)$ is homogeneous of degree $-4$ in the $\e$.

From the holographic calculations, e.g.\ \cite[Eq.~(1.10)]{Martelli:2011fu}, we expect that the leading order squashed sphere free energy relates to the round case via 
\be
\label{eq:squashedsphereleadingorderfreeenergy}
	\cF^\squb_L(\Delta, N) = \frac{(\squb + \squb^{-1})^2 }{4}\, \cF_L (\Delta, N) =   \frac{\sqrt{2} \pi}3\, \frac{(\squb+\squb^{-1})^2\, N^{3/2}}{\sqrt{C_X (\Delta)}}\ ,
\ee
such that the limit $\squb=1$ one recovers the round sphere result.~\footnote{Note that we have introduced a slightly different notation for the free energy with respect to the introductory section, such that $\cF^\squb$ here agrees with $\cF_{S^3_\squb}$ in \eqref{eq:deffreenergy} strictly in the large $N$ limit.}

We then find a natural generalization of the integral representation for the squashed sphere (perturbative) partition function,
\be
	Z_L^\text{pert} (\Delta, \squb, N) = \int \mathd \tilde \mu\, \exp \left(\frac1{4 \pi^2}\, N^{X, \text{mes.}}_{0, 0}  (\tilde \mu, \Delta) - (\squb+\squb^{-1})\,  N\, \tilde \mu \right)\ ,
\ee
leading to
\be
\label{eq:leadingordersquashedspherepartitionfunctionand freenergy}
	- \log Z_L^\text{pert} (\Delta, \squb, N) \stackrel{\text{saddle pt.}}{\approx}   \frac{\sqrt{2} \pi}3\, \frac{(\squb+\squb^{-1})^2\, N^{3/2}}{\sqrt{C_X (\Delta)}}\ ,
\ee
as expected holographically.

\subsubsection*{Conclusion}
We can thus generalize \eqref{eq:mainconjecture-pert-round} to include squashing:
\be
\label{eq:mainconjecture-pert-squashed}
	Z^\text{pert}_L (\Delta, \squb, N_{\rm M2}) = \int \mathd \tilde \mu\, \exp \left(  F^\text{top,pert}_X (\tilde \mu, \frac{(\squb + \squb^{-1})}2\, \Delta; 2 \pi)-  (\squb+\squb^{-1})\, N_{\rm M2}\, \tilde \mu \right)\ .
\ee
However, note that at this point we have not yet looked at the subleading orders in $F_X^\text{top,pert}$, which naturally also depend on an additional refinement parameter $\mathfrak{b}$. The next section serves as a bridge to this discussion, where we eventually conclude that the squashing parameter $\squb$ introduced here is to be identified with the refinement parameter $\mathfrak{b}$.

%%%%%%%%%%%%%%%%%%%%%%%%%%%%%%%%%%%%
\section{Subleading orders in \texorpdfstring{$N$}{N}: refinement from 4d HD supergravity}
\label{sec:3}
Having clarified the relation between the flux $N_{\rm M2}$ and the conjugate variables $\lambda$, we now address another crucial aspect of equivariant topological strings: the possibility of incorporating refinement, defined in \cite{Iqbal:2007ii,Huang:2010kf,Krefl:2010fm} for the Gromov--Witten invariants. The concept of refinement is closely related to the existence of a toric action on non-compact Calabi--Yau manifolds.~\footnote{This explains why refinement is not relevant for the case of vanishing fluxes and the OSV conjecture, as detailed in \cite[Section~7]{Cassia:2025aus}.} Fundamentally, the refinement arises from the correspondence between topological strings and Nekrasov partition functions \cite{Nekrasov:2002qd}, through the so called geometric engineering \cite{Katz:1996fh}. This partition function incorporates two independent equivariant parameters—one for each copy of the complex plane that describes the 4d/5d Omega background of Nekrasov--Okounkov \cite{Nekrasov:2003rj}. One of these parameters corresponds to the string coupling constant, while the other, referred to as the refinement parameter, serves as an additional expansion coefficient.

The refined equivariant topological string partition function, as detailed in the recent exposition \cite[Section~2]{Alexandrov:2023wdj}, has the following formal expansion:
\be
\label{eq:topostringrefinement}
    F^\text{top} (\lambda, \epsilon; g_s, \mathfrak{b}) = \sum_{\mathfrak{g} = 0}^\infty \sum_{n = 0}^\mathfrak{g} (-1)^n g_s^{2 (\mathfrak{g}- 1)}\, (\mathfrak{b} + \mathfrak{b}^{-1})^{2 n}  F^\text{top}_{\mathfrak{g}, n} (\lambda, \epsilon)\ ,
\ee
where the refinement parameter $\mathfrak{b}$ used here corresponds to the one in \cite{Alexandrov:2023wdj}, after a rescaling by $\mathi=\sqrt{-1}$, so that the unrefined limit corresponds to $\mathfrak{b}=\sqrt{-1}$.
%where in some cases the refined Gromov--Witten invariants can be explicitly calculated.
We are however specifically interested in the unique genus-one refined constant maps term, which we denote by $M_{1, 1, 0}$ (more generally $M_{\mathfrak{g}, n = 0, \beta} = N_{\mathfrak{g}, \beta}$),
\be
	 F^\text{top,pert}_{\mathfrak{g} = 1, n = 1} (\lambda, \epsilon; \mathfrak{b}) = - (\mathfrak{b} + \mathfrak{b}^{-1})^2\,  M_{1, 1, 0} (\lambda, \epsilon)\ .
\ee

An apparent caveat of the refinement is the lack of a clear geometric origin for the additional parameter at the level of the CY manifold $X$. This particularly affects the constant maps terms, which lack a general definition through the refined topological vertex of \cite{Iqbal:2007ii}. From the present perspective, however, we are able to infer the refined genus-one constant maps contribution based on the conjectured relation between equivariant topological strings and higher derivative supergravity, illustrated in Fig. \ref{fig:1}.

In particular, the chain of proposed relations enables us to consider the effective 4d supergravity resulting from the dimensional reduction of 11d supergravity on the Sasakian manifold $L$. Remarkably, even without knowing the complete higher derivative action in 11d, 4d supergravity can be analyzed in the superconformal off-shell formalism as initially argued in \cite{Bobev:2020egg,Bobev:2021oku}, giving us a significant advantage. Conceptually, off-shell supergravity in 4d should be thought of as being complementary to the off-shell internal geometry notion, which we presented in \cite[Section~6]{Cassia:2025aus}. This means that we impose supersymmetry on a given background without relying on any equations of motion, thus automatically incorporating supersymmetric fluctuations of the fields around the actual solution. This in turn allows us to make use of the extensive off-shell supergravity discussion that we turn to next in order to find the dependence on the refinement parameter $\mathfrak{b}$. 

In conclusion, as we show in the course of this section, we are able to conjecture the form of the refined constant maps term $M_{1, 1, 0}$ \emph{after} imposing the mesonic twist \eqref{eq:mesonictwist} and the universal parametrization \eqref{eq:univparametrization}, namely
\be
\label{eq:g1equimapref}
 M^{X, \text{mes.}}_{1, 1, 0} (\lam^i_\text{mes.} =\tilde \mu,\e)
 = \frac{\tilde \mu}{24} \sum_{i<j<k} \frac{\partial^3\BV_X(\lam,\e)}{\partial\lam^i\partial\lam^j\partial\lam^k}\Big|_{\lam=0}\,.
% = \frac{\tilde \mu}{24}\oint_\text{JK} \prod_{a=1}^r\frac{\mathd\phi_a}{2\pi\mathi}
% \frac{\sum_{i<j<k}x_ix_j x_k}{\prod_{i=1}^N x_i}\ .
\ee
Note that this is \emph{not} the definition of the refined contribution, but simply the answer that we justify based on the relation with the squashed three-sphere partition function and the respective asymptotic AdS$_4$ background in presence of HD corrections. As expected from the symmetry structure of the topological string expansion, $ M_{1, 1, 0}$ is in close similarity to $ N_{1, 0}$, but lacks a clear geometric origin.

\subsection{4d \texorpdfstring{$\cN=2$}{N=2} off-shell supergravity action}
\label{sec:intermezzo}
We now shift focus to 4d $\cN=2$ supergravity, in order to fully exploit the relations illustrated in Figure~\ref{fig:1}. In particular, we will show that the 4d perspective allows us to rederive the leading order results discussed so far, from the so-called gravitational block viewpoint, as presented in \cite{Hosseini:2019iad}. Furthermore, by leveraging the full power of conformal supergravity, as outlined in \cite{Hristov:2021qsw, Hristov:2024cgj}, we can generalize these results to higher-derivative supergravity. It is worth noting that the ideas presented in \cite{Hosseini:2019iad, Hristov:2021qsw} have recently been placed on a more solid mathematical foundation through a series of papers, see \cite{BenettiGenolini:2023ndb, BenettiGenolini:2024lbj} and references therein. These works utilize the Atiyah--Bott--Berline--Vergne localization theorem in a non-compact setting, offering a rigorous framework for the calculations. In this regard, we are once again employing localization as our primary computational tool, in much the same way as the topological string results.

Here we summarize the main content of \cite{Hristov:2021qsw,Hristov:2022plc,Hristov:2024cgj}, with a focus on key aspects of the 4d $\cN=2$ superconformal formalism, see \cite{Lauria:2020rhc} for a review (we use Euclidean signature, following \cite{deWit:2017cle}). The bosonic symmetry algebra consists of general coordinate transformations, local Lorentz transformations, dilatations, special conformal transformations, and $\SO(1,1) \times \SU(2)$ R-symmetries. On the fermionic side, we have supersymmetry and special supersymmetry transformations. The gauge fields corresponding to general coordinate transformations, dilatations, R-symmetry, and $Q$-supersymmetry are treated as independent fields, while the remaining gauge fields are composite. The explicit matter content for a given model depends on the direct dimensional reduction from the 11d Lagrangian and is not known for an arbitrary internal manifold if we insist on keeping all internal isometries in the compactification ansatz. A notable exception to this general problem is the maximally supersymmetric case of M2-branes on $\BC^4$, where we can keep all four internal $U(1)$'s via a reduction from maximal supergravity to the so called gauged STU model, which serves as a main example for explicit calculations. 

We can write a general expression (to infinite order) for the towers of possible higher derivative invariants and attempt to evaluate these corrections on the supersymmetric backgrounds relevant here. The higher derivative Lagrangian is naturally defined by the generalization of the holomorphic prepotential briefly introduced in our companion paper, \cite[Section~7]{Cassia:2025aus}, now including the double expansions related to the so-called $\mathbb{W}$ and $\mathbb{T}$ invariants,~\footnote{In this subsection and the next one, we use the standard notations for the supergravity prepotential, $F$, and for the complex scalars, $X^I$, not to be confused with the topological string free energy and the underlying manifold, respectively. We believe the distinction is evident from the way the $X$'s appear in the equations, and apologize to the reader for the discomfort.}
\be
\label{eq:HFprepotential}
	F (X; A_\mathbb{W}, A_\mathbb{T}) = \sum_{m, n = 0}^\infty F^{(m,n)} (X)\, (A_\mathbb{W})^m\, (A_\mathbb{T})^n\ ,  
\ee
with each $F^{(m,n)} (X)$ a homogeneous function of degree $2 (1- m-n)$, giving rise to a supersymmetric invariant with $2 (1+m+n)$ number of derivatives. The scalars $X^I$ parametrize the vector multiplets and can be directly related to the equivariant parameters $\epsilon_i$ (see below), while the scalars $A_\mathbb{W, T}$ are composite scalars setting the higher derivative expansion. Additionally, the gauging of the R-symmetry is parametrized by the constant Fayet--Iliopoulos parameters $g_I$, see \cite{Hristov:2021qsw,Hristov:2022plc} for all the explicit details. The gravitational coupling constant, $G_{(4)}$, is hidden inside the coefficients $F^{(m,n)}$, which accommodate arbitrary constant factors. For example, in the two-derivative theory, in the standard conventions we have
\be
\label{eq:2derprepotentialconvention}
	F (X; A_\mathbb{W} = 0, A_\mathbb{T} = 0) =  F^{(0,0)} (X) = \frac{1}{8 \pi\, G_{(4)}}\, F^{2 \partial} (X)\ , 
\ee
with $F^{2 \partial}$ a homogeneous function of degree $2$.

Referring to \cite{Hristov:2022lcw} for more details, the holographic comparison with ABJM at $k=1$ and finite $N$ leads to the following conjecture for the explicit prepotential of the gauged STU model, at all derivative order:
\be
\label{eq:HD-STU}
	F_{\rm STU}(X; A_\mathbb{W}, A_\mathbb{T}) =-\mathi\, \frac{\sqrt{2 X^0 X^1 X^2 X^3}}{3 \pi}\, \,  \left(N - \frac{1}{24} -\frac13\, \frac{  k_{\mathbb{W}} (X) A_{\mathbb{W}} + k_{\mathbb{T}} (X) A_{\mathbb{T}}}{64\, X^0 X^1 X^2 X^3} \right)^{3/2}\ ,
\ee
where $k_{\mathbb{W}} (X)$ and $k_{\mathbb{T}} (X)$ are homogeneous functions of degree $2$ that can be found in \cite{Hristov:2022lcw}, see ft. \ref{ft:match}, and the gauging parameters are all equal, $g_0 = g_1 = g_2 = g_3 = 1$.

Most importantly, it turns out that we can consider the gravitational analog of the Nekrasov background in AdS$_4$ and derive a general infinite-order expression for the (off-shell in this case) supergravity action, as discussed in \cite{Hristov:2022plc}. By employing the supergravity localization argument of \cite{BenettiGenolini:2023ndb}, we are able to evaluate the action on all BPS backgrounds with isolated fixed points w.r.t.\ some {\it toric} action, as shown in \cite{Hristov:2021qsw,Hristov:2024cgj}. The general form of the supergravity action on a supersymmetric background $M_4$, defined via \eqref{eq:HFprepotential}, is given by,  \cite{Hristov:2021qsw, Hristov:2024cgj}
\be
\label{eq:conj1}
\begin{aligned}
	\cI_{\partial M_4} =&\,  \sum_{\sigma \in M_4^{\rm fixed}} \cB (X_{(\sigma)}, b_1^{(\sigma)}, b_2^{(\sigma)})\ , \\
	\cB(X, b_1, b_2) :=&\, \frac{4 \i \pi^2}{b_1 b_2}\, F( X; (b_1-b_2)^2, (b_1+b_2)^2)\ ,
\end{aligned}
\ee
subject to the so-called gluing rule: the identification of the $X^I_{(\sigma)}$ (dubbed Coulomb branch parameters) and $b_{1,2}^{(\sigma)}$ (the equivariant parameters from 4d perspective) at the different fixed points, together with an additional linear constraint $\lambda^{M_4} (X^I, b_{1,2}) = 0$. In the present case, we are only going to focus on the supergravity dual of the squashed sphere partition function, meaning that the explicit calculation of \cite{Hristov:2022plc} is enough, without the need for invoking the more general gluing rules in \cite{Hristov:2021qsw,Hristov:2024cgj}.

\subsection{\texorpdfstring{$\Omega\mathbb{H}^4$}{OH4} background}
\label{sec:OmegaBack}
Let us now look concretely at the 4d asymptotically (Euclidean) AdS$_4$ background, which is holographically dual to the partition function on a squashed $S^3$, \cite{Martelli:2011fu}. Geometrically, this background corresponds to the maximally symmetric vacuum, Euclidean AdS$_4$ (denoted as $\mathbb{H}^4$), with a round three-sphere boundary. To explicitly break the full symmetry group down to $U(1) \times U(1)$, we introduce either a purely self-dual or purely anti-self-dual background tensor field, encoding the squashing parameter $\squb$. The resulting space is referred to as the Omega background, denoted by $\Omega \mathbb{H}^4$. For further details, see \cite{Hristov:2022plc}, which provides a detailed discussion of Omega backgrounds in the context of supergravity. Specifically, it was shown in \cite{Hristov:2022plc} that the gauged supergravity generalization of the usual $\Omega \BR^4$ background, introduced by Nekrasov and Okounkov \cite{Nekrasov:2003rj}, is precisely the space studied in \cite{Martelli:2011fu}.

The $\Omega \mathbb{H}^4$ background exhibits a single fixed point, at the origin, under the aforementioned symmetry group $U(1)^2$. The identification of the equivariant parameters in terms of the squashing $\squb$ was performed explicitly in \cite{Hristov:2022plc}, with the natural choice
\be
	\frac{b_2}{b_1} = \squb^2\ , \qquad \squb = \sqrt{\frac{b_2}{b_1}}\ ,
\ee
where we assumed $\squb > 0$ without loss of generality. Note that the final answer is invariant under $\squb \rightarrow 1/\squb$, such that we can choose the ratio of $b_1$ and $b_2$ in any order. This reflects the fact that the overall normalization of the torus action is inconsequential in supergravity.

We also have the Coulomb branch parameter identification,
\be
\label{eq:s3gluing}
	X^I = \frac{(b_1 + b_2)}2\, \varphi^I = b_1\, \frac{(1+\squb^2)}2\, \varphi^I\ , 
\ee
together with the supersymmetric constraint
\be
\label{eq:s3susyconstrsugra}
	g_I \varphi^I = 2\ . 
\ee

\subsection*{Two-derivative result}
Plugging these identifications in the two-derivative prepotential, we find the following supergravity action on the $\Omega \mathbb{H}^4$ background,~\footnote{We use the short-hand notation $\partial\, \Omega \mathbb{H}^4 = S^3_\squb$, even though strictly speaking this is only true after a conformal transformation of the boundary.}
\be 
	\cI^{2 \partial}_{S^3_\squb} (\varphi, \squb) = \frac{4 \mathi \pi^2}{b_1 b_2}\, F^{(0,0)} \left(\frac{(b_1 + b_2)}2\, \varphi \right) = \frac{\pi\mathi}{8\, G_{(4)}}\,  (\squb+\squb^{-1})^2\, F^{2 \partial} (\varphi)\ ,
\ee
using the homogeneity of the prepotential and the previous identities. This is the supergravity prediction for the leading order free energy of the squashed $S^3$ partition function, which should precisely agree with the topological string calculation from the previous section, \eqref{eq:leadingordersquashedspherepartitionfunctionand freenergy}.

Let us for simplicity assume $g_I = 1, \forall I$, since we are always allowed to rescale the variables $\varphi^I$. Comparing with the expressions in the previous section and the supersymmetric constraints, \eqref{eq:Deltaconstr} and \eqref{eq:s3susyconstrsugra}, we conclude that we need to identify
\be
	\varphi^I\, ``=" \,\Delta_i\ ,
\ee
where we need to appropriately identify the indices $I$ with $i$ (and neglect the difference in positioning, which is a matter of convention). 
We can further use the holographic identifications,
\be
\label{eq:2derholograpaphicidentifications}
	\frac1{4\, G_{(4)}} = \frac{\sqrt{2}}{3}\, N^{3/2}\ , \qquad F^{2 \partial} (\varphi) = - \frac{2\mathi}{\sqrt{C_X (\Delta)}}\ ,
\ee
such that the leading order AdS/CFT result gives
\be
	\cI^{2 \partial}_{S^3_\squb} (\varphi, \squb) = \cF^\squb (\Delta, N)\ ,
\ee
where $\cF^\squb$ specifically denotes the large $N$ answer for the dual free energy, see \eqref{eq:squashedsphereleadingorderfreeenergy}.

In particular, for the STU prepotential, \eqref{eq:HD-STU}, at large $N$, we find
\be
    F^{(0,0)}_{\rm STU} = -\mathi\, \frac{\sqrt{2 X^0 X^1 X^2 X^3}}{3 \pi}\, \,  N^{3/2} = \frac{\sqrt{2}\, N^{3/2}}{6 \pi}\, F^{2 \partial}_{\rm STU}\ ,
\ee
such that
\be
    \cF^\squb_{\rm ABJM} (\Delta, N) = \frac{\pi\, N^{3/2}}3\, (\squb+\squb^{-1})^2\, \sqrt{2 \Delta_1 \Delta_2 \Delta_3 \Delta_4}\ ,
\ee
in agreement with \eqref{eq:squashedsphereleadingorderfreeenergy}, given that $C_{\BC^4}(\e) = 1/\left( \e_1 \e_2 \e_3 \e_4 \right)$.

\subsection*{Higher-derivative prediction}
A direct evaluation of the off-shell supergravity action of the $\Omega \mathbb{H}^4$ background was performed in \cite{Hristov:2022plc}, confirming that the two-derivative identifications for this background, \eqref{eq:s3gluing}-\eqref{eq:s3susyconstrsugra}, continue to hold at the level of the full higher-derivative expression in \eqref{eq:conj1}. This leads to
\be 
	\cI_{S^3_\squb} (\varphi, \squb) = \frac{4 \mathi \pi^2}{\squb^2}\, F \left((1-\squb^2)\, \frac{\varphi}2; (1-\squb^2)^2, (1+\squb^2)^2 \right)\ ,
\ee
where we used the homogeneity properties of the functions $F^{(m,n)} (X^I)$ to simplify the above expression.  

Using the higher-derivative STU model, \eqref{eq:HD-STU}, which is dual to ABJM at level $k=1$ and relates to the topological string on $\BC^4$, we arrive at the prediction
\be 
\hspace{-5pt}
\label{eq:HDSTUresult}
	\cI^{\rm STU}_{S^3_\squb} (\varphi, \squb) = \pi\, (\squb+\squb^{-1})^2\, \frac{\sqrt{2 \varphi^0 \varphi^1 \varphi^2 \varphi^3}}{3}\, \,  \left(N - \frac{1}{24} -\frac{(\squb-\squb^{-1})^2\,  k_{\mathbb{W}} (\varphi)  + (\squb+\squb^{-1})^2\, k_{\mathbb{T}} (\varphi)}{48\, \varphi^0 \varphi^1 \varphi^2 \varphi^3} \right)^{3/2}\ .
\ee

\subsection{Refinement identification}
\label{sec:3.1}
After the digression into the effective four-dimensional supergravity formalism, we are now ready to come back to the topological string side of the calculation. Following the chain of identifications we used for the evaluation of $N^{X, \text{mes.}}_{0,0}$, we can analogously simplify the (unrefined) genus-one contribution,
\be
	N^{X, \text{mes.}}_{1, 0} (\tilde \mu, \e) = \frac{(\squb+\squb^{-1})\,\tilde \mu}{24}\, k_2 (\e)\, C_X (\e) =  \frac{\tilde \mu}{6\, (\squb+\squb^{-1})}\, k_2 (\Delta)\, C_X (\Delta) \ ,
\ee
where we used \eqref{eq:squashedepsilontob} and the definitions \eqref{eq:shorthandc2andc3}, along with the fact that the product $k_n(\e)C_X(\e)$ is homogeneous of degree $n-4$ in $\e$. Additionally our conjecture, \eqref{eq:g1equimapref}, for the refined contribution gives
\be
	M^{X, \text{mes.}}_{1, 1,0}  (\tilde \mu, \e) = \frac{\tilde \mu}{24}\, k_3(\e)\, C_X (\e) = \frac{\tilde \mu}{12\, (\squb+\squb^{-1})}\, k_3(\Delta)\, C_X (\Delta)\ ,
\ee
where we again used \eqref{eq:shorthandc2andc3}.

Now we want to evaluate $Z_L^\text{pert}$ from \eqref{eq:mainconjecture-pert-squashed} in the presence of both genus-zero and the refined genus-one terms,
\be
%\hspace{-15pt}
\label{eq:transfsubleading}
\begin{aligned}
	Z_L^\text{pert} (\Delta, \squb, N_{\rm M2}) &= \int \mathd \tilde \mu\, \exp \left( \frac1{4 \pi^2}\, N^{X, \text{mes.}}_{0, 0}+ N^{X, \text{mes.}}_{1, 0} -  (\mathfrak{b} + \mathfrak{b}^{-1})^2 M^{X, \text{mes.}}_{1, 1,0}  -  (\squb+\squb^{-1})\, N_{\rm M2}\, \tilde \mu  \right)\\
&= \int \mathd \tilde \mu\, \exp \left(\frac1{4 \pi^2}\, \frac{8\, C_X (\Delta)}{3\, (\squb+\squb^{-1})}\, \tilde \mu^3 - (\squb+\squb^{-1})\, \left(N_{\rm M2} - B_X (\Delta, \squb)\, \right)\, \tilde \mu \right)\ ,
\end{aligned}
\ee
where we defined
\be
	B_X (\Delta, \squb):=\frac1{12}\, \frac{C_X (\Delta)}{(\squb+\squb^{-1})^2}\, \left(2\, k_2 (\Delta) - (\mathfrak{b} + \mathfrak{b}^{-1})^2\, k_3(\Delta) \right)\ .
\ee

The saddle point evaluation of the above integral results in,~\footnote{We are now evaluating the expression exactly in $N$, such that the saddle point approximation is precisely the free energy $\cF_{S^3_\squb}$ defined in \eqref{eq:deffreenergy}.}
\be
\label{eq:freenergysquashedsphere}
	- \log Z_L^\text{pert} (\Delta, \squb, N_{\rm M2})  \stackrel{\text{saddle pt.}}{\approx} \frac{\sqrt{2} \pi\, (\squb+\squb^{-1})^2}{3\, \sqrt{C_X (\Delta)}} \left(N_{\rm M2}- B_X (\Delta, \squb) \right)^{3/2}  = \cF_{S^3_\squb, L} (\Delta, N_{\rm M2})\ .
\ee
Note that at this stage, even though we went ahead of the logic by using the conjecture for $M_{1,1,0}$, \eqref{eq:g1equimapref}, in the definition of $B_X$, we could still simply regard the final form of $\cF_{S^3_\squb}$ here as a formal expression. The precise answer for the refinement, which ultimately fixes the form of $B_X$ will now be tested against the higher-derivative prediction for the STU model described in the previous subsection.

\subsubsection*{Match with STU supergravity}
We have now derived an expression for the free energy of the M2-brane theory, via the topological string picture, including finite $N_{\rm M2}$ corrections, where we have also included the exact M2-brane charge, see \cite[Appendix~A]{Cassia:2025aus}. We now invoke holography to make full use of the effective HD supergravity results presented in the preceding subsections. Recall that the STU result in \eqref{eq:HDSTUresult} is expected to come from eleven-dimensional supergravity compactified on $S^7$, dual to ABJM theory at $k=1$,
\be
	\cF_{S^3_\squb}^{\rm ABJM} (\e, N_{\rm M2}) \stackrel{!}{=}  \cI^{\rm STU}_{S^3_\squb} (\squb, \varphi^I)\ ,
\ee
where the underlying toric manifold is $X = C(S^7) = \BC^4$. In order to avoid repetition, we refer the reader to Section~\ref{sec:ABJM} for the equivariant volume in this case and a more detailed discussion of the corresponding values of $k_2, k_3$ and $N_{\rm M2}$. Here we simply cite the results,
\be
    C (\Delta) = \frac{1}{\prod_{i=1}^4 \Delta_i}\ , \quad \chi = 1\ , \quad k_2 (\Delta) = \sum_{i < j} \Delta_i \Delta_j\ , \quad k_3 (\Delta) = \sum_{i<j<k} \Delta_i \Delta_j \Delta_k\ .
\ee

After a careful examination of these expressions and the comparison with \eqref{eq:HDSTUresult}, we arrive at the simple relation between the refinement parameter $\mathfrak{b}$ and the squashing parameter $\squb$,~\footnote{\label{ft:match}In order to derive the exact match, we note the useful relations with the quantities $k_{\mathbb{T},\mathbb{W}}$ in \cite{Hristov:2022lcw}: \be  k_\mathbb{T}^\text{there} (\Delta) = 2\, k_2^\text{here} (\Delta) - 4\, k_3^\text{here} (\Delta)\ , \qquad  k_\mathbb{W}^\text{there} (\Delta) = -2\, k_2^\text{here} (\Delta)\ ,   \ee where in the first relation we used the constraint $\sum_i \Delta_i = 2$.}
\be
\label{eq:reftosquash}
	\mathfrak{b} = \squb\ ,
\ee
which is only possible after we have used our conjecture for $M_{1,1,0}$ as in \eqref{eq:g1equimapref}. Note that for the ease of presentation we have inverted here the actual order of our calculations, but in fact the conjecture for the refined genus-one expression is precisely fitted to account for the present holographic match. Therefore we can consider ABJM theory at level $k=1$ as a tuning point, allowing us to fix uniquely the refined topological string expression. Once this is done, all other examples of manifolds $X$ give predictions for the squashed sphere partition functions of the corresponding M2-brane theories.

The identification \eqref{eq:reftosquash} follows directly from the observation that $g_s$ and $\mathfrak{b}$ are the Nekrasov parameters in the refined topological string expansion \eqref{eq:topostringrefinement}, while $b_{1,2}$ denote the equivariant parameters of the $\Omega \mathbb{H}^4$ background. In our framework, $g_s$ acts merely as a normalization constant, and in supergravity only the ratio $b_1/b_2$ is physically relevant. Consequently, \eqref{eq:reftosquash} matches the two independent parameters, $\mathfrak{b}$ and $\squb$, on the respective sides. It is important to remark, however, that the unrefined limit on the topological string side does \emph{not} correspond to the round sphere limit on the field theory side.

\subsubsection*{Conclusion}
We have now justified the proposed refinement of the equivariant topological string genus-one constant maps term, and have additionally derived the relation between the refinement parameter and the squashing as in \eqref{eq:reftosquash}. Using this, we can now simplify the expression for $B_X$,
\be
	 B_X (\Delta, \squb) = \frac{C_{X} (\Delta)}{12}\, \left( \frac{2\, k_2 (\Delta)}{(\squb+\squb^{-1})^2} - k_3 (\Delta) \right)\ .
\ee
such that the saddle point free energy, $\cF_{S^3_\squb}$, is still given by \eqref{eq:freenergysquashedsphere}. As demonstrated above via holography, this is precisely recovered by the higher-derivative supergravity action. This justifies relation \eqref{eq:HDsugractiongeneral}, where we note that the saddle-point evaluation of the integral \eqref{eq:transfsubleading} precisely recovers the leading exponent in the asymptotic expansion of the Airy function, \eqref{eq:airyasympt}.

What now remains to be done is the exact evaluation of \eqref{eq:transfsubleading}, namely
\be
\label{eq:finalintegral}
    Z_L^\text{pert} (\Delta, \squb, N_{\rm M2}) = \int \mathd \tilde \mu\, \exp \left(\frac1{4 \pi^2}\, \frac{8\, C_X (\Delta)}{3\, (\squb+\squb^{-1})}\, \tilde \mu^3 - (\squb+\squb^{-1})\, \left(N_{\rm M2} - B_X (\Delta, \squb)\, \right)\, \tilde \mu \right)\ ,
\ee
using the simplified expression for $ B_X (\Delta, \squb)$ above.

%%%%%%%%%%%%%%%%%%%%%%%%%%%%%%%%%%%%%%%%%%%%%%%%%%
\section{Perturbatively exact in \texorpdfstring{$N$}{N}}
\label{sec:5}
Putting everything together, we can evaluate \eqref{eq:finalintegral} using the knowledge of the integral representation of the Airy function in \eqref{eq:airyasympt}. Since we omitted the higher-genus constant maps terms, which carry no dependence on $\tilde \mu$, as well as the non-perturbative completion of the topological string partition function, the answer holds up to constant prefactors and instanton corrections:
\be
\label{eq:fullanswersec5}
	Z_{S^3_\squb}^\text{pert}  \simeq {\rm Ai} \Big[ \left( \frac{2\, C_{X} (\Delta)}{\pi^2\, (\squb+\squb^{-1})^4}\right)^{-1/3} \left(N_{\rm M2}- \frac{ C_{X} (\Delta)}{12}\, \left(\frac{2\, k_2 (\Delta)}{(\squb+\squb^{-1})^2} - k_3 (\Delta) \right) \right) \Big]\ ,
\ee
as advertised in the introduction. This gives a prediction for the squashed sphere partition function of an infinite set of M2-branes models on the tip of toric Sasakian manifolds. We remind the reader that the number of mesonic $\Delta$ parameters depends on how $X$ is realized as a symplectic quotient. They are subject to the constraint \eqref{eq:holocond2}, but nevertheless in general correspond to an overparametrization of the independent supersymmetric deformations due to the existence of flat directions in the resulting free energy $\cF_{S^3_\squb}$, \eqref{eq:deffreenergy}, see \cite{Jafferis:2011zi}. We can thus view the additional mesonic constraint, \eqref{eq:mesonicconstraint}, as a particular prescription for lifting the flat directions, which precisely agrees with the choice in \cite{Jafferis:2011zi}.
Moreover, it is worth noting that while $C_X(\Delta)$ is uniquely defined independently of the choice of chamber for the symplectic quotient, quantities like $k_2(\Delta)$, $k_3(\Delta)$ and $\chi(X)$ that appear in the subleading corrections in \eqref{eq:fullanswersec5}, can a priori be different in different chambers.

Next, we turn to discuss explicitly the evaluation of the above general formula for a number of prominent M2-brane models.

\subsection{ABJM theory}
\label{sec:ABJM}
We start with ABJM theory at level $k$, \cite{Aharony:2008ug}, originally preserving $\cN=6$ supersymmetry ($\cN=8$ for $k=1,2$) before introducing arbitrary R-charges which break supersymmetry down to $\cN=2$. In this case $L = S^7/\BZ_k$, such that $X=\BC^4/\BZ_k$. This space was considered in \cite[Section~3.1]{Cassia:2025aus}. Note that for $k=1$ we have a regular space, but at $k>1$ in principle we need to resolve the conical singularity first, evaluate the equivariant volume and only then blow down the additional cycles to get back to the cone. Since this case is rather straightforward, we immediately give the results in terms the four effective K\"ahler parameters $\lam^i$ and four equivariant parameters $\e_i$  after the mesonic twist,
\be
	\BV_{\BC^4/\BZ_k} = \frac{\mathe^{\lam^i \e_i}}{k\, \prod_{i=1}^4 \e_i}\ ,
    \hspace{30pt}
    \text{such that}
    \hspace{30pt}
	C_{\BC^4/\BZ_k} (\e) = \frac1{k\, \prod_{i=1}^4 \e_i}\ .
\ee
The evaluation of the terms $k_2$ and $k_3$ also follows from \eqref{eq:shorthandc2andc3},
\be
	k_2 (\e) = \sum_{i < j} \e_i \e_j\ , \qquad k_3 (\e) = \sum_{i < j < k} \e_i \e_j \e_k\ .
\ee
On the other hand, it is clear that the evaluation of the Euler number cannot originate directly from taking derivatives of $\BV$ via \eqref{eq:exactchargeviaEulerchar}, which would give $1/k$. Instead, we can compute the Euler characteristic by going to the resolved space. After the resolution, we can simply count the number of fixed-points of the torus action, and we obtain 
\be
	\chi (\BC^4/\BZ_k) = k\ , \quad \Rightarrow \quad N_{\rm M2} = N - \frac{k}{24}\  . 
\ee

We now have all ingredients to plug inside of \eqref{eq:fullanswersec5},
\be
\hspace{-20pt}
\label{eq:fullanswerABJM}
	Z_\text{ABJM}^\text{pert}  \simeq {\rm Ai} \Big[ \left( \frac{2}{\pi^2 (\squb+\squb^{-1})^4\, k\, \prod_i \Delta_i}\right)^{-1/3} \left(N_{\rm M2}- \frac{1}{12 k\, \prod_i \Delta_i}\, \left(\frac{2\,  \sum_{i < j} \Delta_i \Delta_j}{(\squb+\squb^{-1})^2} - \sum_{i < j < k} \Delta_i \Delta_j \Delta_k \right) \right) \Big]\ ,
\ee
This result is in exact agreement with the prediction in \cite{Hristov:2022lcw} from HD supergravity and the field theory conjecture of \cite{Bobev:2022eus,Bobev:2025ltz}, noting that the last term proportional to $k_3 (\Delta)$ has been conjectured based on this agreement. The remaining terms have instead been rigorously derived from the topological string framework. It can be compared and successfully shown to match with a number of exact localization results on the field theory side, see \cite{Herzog:2010hf,Fuji:2011km,Marino:2011eh,Nosaka:2015iiw,Hatsuda:2016uqa,Chester:2021gdw,Geukens:2024zmt,Kubo:2024qhq} and references therein.

The geometric part of the above results can be straightforwardly extended to the case of fractional brane charges, corresponding to ABJ theory \cite{Aharony:2008gk}, since the underlying manifold $X$ remains unchanged. The only modification is an additional shift in $N_{\rm M2}$, as discussed in \cite{Bergman:2009zh}. On the other hand, we cannot exclude the possibility that fractional branes contribute to the refinement of the genus-one term, responsible for the $k_3$ correction identified above. We leave a more systematic study of ABJ theory for future investigation; see \cite{Matsumoto:2013nya,Honda:2014npa} and references therein for exact results on the field theory side.

\subsection{ADHM theory and circular quivers}
The so called ADHM theory, \cite{Atiyah:1978ri,deBoer:1996mp,Porrati:1996xi}, originally preserves $\cN=4$ supersymmetry and comes from M2-branes with a rather similar transverse geometry: $\BC^2 \times \BC^2/\BZ_{N_f}$.The value of $N_f$ corresponds to the $SU(N_f)$ flavor symmetry of the corresponding quiver. From our perspective it is convenient to consider the ADHM together with other circular quivers preserving $\cN = 4$ and $\cN = 3$ supersymmetry, see \cite{PeterBKronheimer:1990zmj,deBoer:1996mp,Porrati:1996xi,Imamura:2008nn,Cremonesi:2016nbo} and references therein. We choose the most general geometry incorporating all these subcases, which corresponds to
\be
X = (\BC^2/\BZ_p \times \BC^2/\BZ_q)/\BZ_k\ ,
\ee 
such that we can recover e.g.\ the ADHM theory by the special choice $p=1, q = N_f, k=1$.

Similarly to the discussion for ABJM above, one in principle needs to resolve the conical singularity in order to derive all quantities entering our general answer, \eqref{eq:fullanswersec5}. Due to the simplicity of the manifold, we first state the results, and turn to discuss the less obvious part of the calculation with an explicit example below. We first find the mesonic equivariant volume,
\be
	C_{X} (\e) =\frac1{k}\,  \frac1{p\, \e_1 \e_2}\, \frac1{q\, \e_3 \e_4} =  \frac1{k\, p\, q\, \prod_i \e_i}\ .
\ee
together with the Euler characteristic\footnote{We compute the Euler characteristic by resolving the singularity to a geometric phase.}
\be
	\chi (X) = k\, p\, q\ , \quad \Rightarrow \quad N_{\rm M2} = N - \frac{k p q}{24}\  . 
\ee
The evaluation of the terms $k_2$ and $k_3$ from \eqref{eq:shorthandc2andc3} gives
\be
\label{eq:k2k3forquotients}
	k_2 (\e) = p^2\, \e_1 \e_2 + q^2\, \e_3 \e_4 + (\e_1+\e_2) (\e_3+\e_4)\ , \quad  k_3 (\e) = p^2\, \e_1 \e_2 (\e_3+\e_4) + q^2\, \e_3 \e_4 (\e_1+\e_2)\ .
\ee
%The reason for the additional factors of $q^2$ and $p^2$ is that in those terms the quantities $k_2$ and $k_3$ compute the Euler characteristic of $\BC^2/\BZ_q$ and $\BC^2/\BZ_p$, respectively.

Given the above results, we can explicitly evaluate \eqref{eq:fullanswersec5} and obtain prediction for the partition functions of the ADHM and the circular quivers. We refrain from repeating the formula, which resembles \eqref{eq:fullanswerABJM} with the additional dependence on $q$ and $p$ as spelled out above. The answer can be compared and successfully matched with the localization results in \cite{Herzog:2010hf,Nosaka:2015iiw,Kubo:2025dot}, which correspond to different limits of our general expression, and to the conjecture in \cite{Bobev:2025ltz} about the ADHM quiver partition function. 

We finish this subsection with a more explicit example of how the extra factors of $q, p$ appear above, using the results in \cite[Section~4.4]{Cassia:2025aus}. We can take the mesonic twist on the $A_2$ geometry, which corresponds to the resolution of $\BC^2/\BZ_3$.  In this case we a priori have four equivariant parameters $\e_{1,\dots,4}$, but the mesonic constraint, \eqref{eq:mesonicconstraint}, simply truncates two of them, $\e_2 = \e_3 = 0$. The equivariant volume (in the geometric chamber $t^1, t^2 > 0$) is given by
\be
 \BV_{A_2}(\lam,\e) = -\frac{\mathe^{3 \lam^3 \e_1-\lam^4(2\e_1-\e_4)}}
 {3\e_1 (2\e_1-\e_4)}
 -\frac{\mathe^{\lam^2(2\e_1-\e_4)-\lam^3(\e_1-2\e_4)}}
 {(\e_1-2\e_4)(2\e_1-\e_4)}
 +\frac{\mathe^{\lam^1(\e_1-2\e_4)+3 \lam^2 \e_4}}
 {3 (\e_1-2\e_4) \e_4} \ ,
\ee
such that
\be
 \BV_{A_2} (\lam_\text{mes.},\e) = \frac{\mathe^{\lam_\text{mes.}^1 \e_1+\lam_\text{mes.}^4 \e_4}}
 {3 \e_1 \e_4} = \mathe^{\e_i \lam^i_\text{mes.}}\, C_{\BC^2/\BZ_3}\ .
\ee
The $k_2$ term and the Euler character are readily evaluated as~\footnote{On two-dimensional manifolds, $k_3 = 0$ identically. Note also that these results are strictly valid in the chosen chamber, which is the only one corresponding to a geometric phase. This subtlety is not present in the other examples we study below.}
\be
	\chi (\BC^2/\BZ_3) = 3 = 9\, \e_1 \e_4\, C_{\BC^2/\BZ_3}\ , \qquad k_2 = \frac{\chi (\BC^2/\BZ_3)}{C_{\BC^2/\BZ_3}} = 9\, \e_1 \e_4 \ .
\ee
For comparison, the same quantities on $\BC^2$ are given by
\be
	\chi (\BC^2) = 1 = \e_1 \e_2\, C_{\BC^2}\ , \qquad k_2 = \frac{\chi (\BC^2)}{C_{\BC^2}} = \e_1 \e_2 \ .
\ee
Taking the product of the two spaces (relabeling $\e_1$ of $A_2$ as $\e_3$), we find
\be
\begin{aligned}
	k_2 (\BC^2 \times \BC^2/\BZ_{3}) &= \e_1 \e_2 + (\e_1+\e_2) (\e_3+\e_4) + 3^2\, \e_3 \e_4\ ,\\
	k_3 (\BC^2 \times \BC^2/\BZ_{3}) &= \e_1 \e_2 (\e_3+\e_4) +9\, \e_3 \e_4 (\e_1+\e_2)\ ,
\end{aligned}
\ee
in agreement with \eqref{eq:k2k3forquotients}.

\subsection{Flavored ABJM}
An example from \cite{Benini:2009qs,Cremonesi:2010ae} is the flavored ABJM model, corresponding to the toric manifold $X = \BC \times \cC$, with $\cC$ being the resolved conifold, i.e.\ the resolution of the singular cone described by the quadratic equation in $\BC^4$:
\be
	z_1 z_3 - z_2 z_4 = 0\ .
\ee 
If we place the flat direction $\BC$ inside of $X$ as the first coordinate (with index $0$ below), the charge vector can be written as
 \be
	Q = \left(0 \quad 1 \quad 1 \quad -1 \quad -1 \right)\ ,
\ee
and the corresponding equivariant volume is 
\be
	\BV (\lambda, \e)=  \frac{\mathe^{\lam^0 \e_0}}{\e_0 (\e_2-\e_1)}\, \left( \frac{\mathe^{\lambda^2 (\e_2-\e_1)+\lambda^3 (\e_3+\e_1) +\lambda^4 (\e_4+\e_1)}}{(\e_3+\e_1) (\e_4+\e_1)} - \frac{\mathe^{\lambda^1 (\e_1-\e_2)+\lambda^3 (\e_3+\e_2) +\lambda^4 (\e_4+\e_2)}}{(\e_3+\e_2) (\e_4+\e_2)} \right)\ .
\ee
We also find
\be
	\chi(X) = \chi(\BC)\chi(\cC) = 2\ , \quad \Rightarrow \quad N_{\rm M2} = N - \frac{1}{12}\  . 
\ee
In this case there is a mesonic twist condition, given by
\be
	\lam_\text{mes.}^1 + \lam_\text{mes.}^2 = \lam_\text{mes.}^3+\lam_\text{mes.}^4\ .
\ee
The mesonic constraint, which can be used to lift the redundancy in the equivariant parameters, simplifies to
\be
\label{eq:conifoldmesonicconstr}
	\e_1 \e_2 = \e_3 \e_4\ .
\ee
Using the mesonic twist, we can write
\be
	\BV_{\BC \times \cC} = \frac{\mathe^{\lambda_\text{mes.}^i \e_i}\, (\e_1+\e_2+\e_3+\e_4)}{\e_0 (\e_1+\e_3) (\e_2+\e_3) (\e_1+\e_4) (\e_2+\e_4)} =: \mathe^{\lambda_\text{mes.}^i \e_i}\, C_{\BC \times \cC}\ ,
\ee
while the evaluation of the terms $k_2$ and $k_3$ from \eqref{eq:shorthandc2andc3} gives
\be
\begin{aligned}
	k_2 (\e) &= \e_1 \e_2 + \e_3 \e_4 +  \sum_{0\leq i < j \leq 4} \e_i \e_j  \ , \\ 
k_3 (\e) &= \e_0 \left(\e_1 \e_2+\e_3 \e_4 + \sum_{1\leq i < j \leq 4} \e_i \e_j \right) +2 \frac{(\e_1+\e_3) (\e_2+\e_3) (\e_1+\e_4) (\e_2+\e_4)}{\e_1 +\e_2+\e_3+\e_4}\ .
\end{aligned}
\ee
We are not aware of any references discussing the field theory calculation of the partition function beyond the leading order in the gauge group rank $N$, which simply matches the calculation for $C_{\BC \times \cC}$, \cite{Jafferis:2011zi}.

\subsection{Other models}
Our general formula can be readily applied to any toric CY four-fold, which can always be thought of as (a resolution of) the cone over a 7d Sasakian space.
%It is impossible, and not instructive, to go on with many examples.
We just mention here a couple of other examples, which have been previously discussed in the literature.

If we consider the cone over $Q^{1,1,1}$, we can immediately observe that the field theory dual is well known due to its quiver description as a different flavoring of the ABJM model, compared to the conifold example above, \cite{Benini:2009qs}. As discussed in \cite{Cassia:2025aus}, the model is defined by the charge matrix
\be
Q =
\begin{pmatrix}
1 & 1 & 0 & 0 & -1 & -1 \\
0 & 0 & 1 & 1 & -1 & -1
\end{pmatrix}
\ee
leading to the equivariant volume
\be
\ba
 \BV_{C (Q^{1,1,1})} (\lam,\e) =&\frac{\mathe^{\lam^2 (\e_2 - \e_1) + \lam^4 (\e_4-\e_3) + \sum_{j=5}^6 \lam^j (\e_j+ \e_1+\e_3)}}{(\e_2-\e_1)(\e_4-\e_3)  \prod_{j=5}^6 (\e_j+ \e_1+\e_3)} +  \frac{\mathe^{\lam^2 (\e_2 - \e_1) + \lam^3 (\e_3-\e_4) + \sum_{j=5}^6 \lam^j (\e_j+ \e_1+\e_4)}}{(\e_2-\e_1)(\e_3-\e_4) \prod_{j=5}^6 (\e_j+ \e_1+\e_4)}\\
 &+ \frac{\mathe^{\lam^1 (\e_1 - \e_2) + \lam^4 (\e_4-\e_3) + \sum_{j=5}^6 \lam^j (\e_j+ \e_2+\e_3)}}{(\e_1-\e_2)(\e_4-\e_3) \prod_{j=5}^6 (\e_j+ \e_2+\e_3)} +  \frac{\mathe^{\lam^1 (\e_1 - \e_2) + \lam^3 (\e_3-\e_4) + \sum_{j=5}^6 \lam^j (\e_j+ \e_2+\e_4)}}{(\e_1-\e_2)(\e_3-\e_4) \prod_{j=5}^6 (\e_j+ \e_2+\e_4)} \ .
\ea
\ee
Finding explicit expressions for $C_X, k_2,$ and $k_3$ is now straightforward, but we refrain from giving them here due to their considerable length (after the mesonic constraint, $C_X$ agrees with the large $N$ evaluation of the field theory, \cite{Jafferis:2011zi}). We only record the answer for the Euler number, giving the M2-brane charge in this case,
\be
	\chi (C (Q^{1,1,1}))=  \sum_{i < j < k < l} \frac{\partial^4 \BV (\lam,  \e)}{\partial \lam^i \partial \lam^j \partial \lam^k \partial \lam^l} \Big|_{\lam = 0} = 4\ , \quad \Rightarrow \quad N_{\rm M2} = N - \frac1{6}\ .
\ee

We can also consider the cone over $M^{1,1,1}$, which enjoys a similar supergravity description, but is far less understood as a gauge theory. The model is defined by the charge vector
\be
 Q = \left(3,3,-2,-2,-2\right)\ ,
\ee
leading to the equivariant volume, \cite{Cassia:2025aus},
\begin{multline}
 \BV_{C (M^{1,1,1})} (\lam, \e) = \frac1{3(\e_2-\e_1)}\,
 \Big( \frac{\mathe^{\lam^2 (\e_2-\e_1)+\lam^3 (\e_3+\tfrac23\e_1)
 +\lam^4 (\e_4+\tfrac23\e_1) + \lam^5 (\e_5+\tfrac23 \e_1)}}
 {(\e_3+\tfrac23 \e_1) (\e_4+\tfrac23 \e_1) (\e_5 + \tfrac23 \e_1)}\\
 - \frac{\mathe^{\lam^1 (\e_1-\e_2)+\lam^3 (\e_3+\tfrac23 \e_2)
 +\lam^4 (\e_4+\tfrac23 \e_2) + \lam^5 (\e_5+\tfrac23 \e_2)}}
 {(\e_3+\tfrac23 \e_2) (\e_4+\tfrac23 \e_2) (\e_5 + \tfrac23 \e_2)} \Big)\ .
\end{multline}
Again, for brevity, we refrain from giving the explicit expressions for $C_X, k_2$, and $k_3$, recording only the brane charge,
\be
	\chi (C (M^{1,1,1}))=  \sum_{i < j < k < l} \frac{\partial^4 \BV (\lam,  \e)}{\partial \lam^i \partial \lam^j \partial \lam^k \partial \lam^l} \Big|_{\lam = 0} = \frac23\ , \quad \Rightarrow \quad N_{\rm M2} = N - \frac1{36}\ .
\ee
We notice however, that in this case, both chambers for the symplectic quotient are non-geometric (i.e.\ orbifold), hence the Euler characteristic is not an integer.

\section{Black holes, twisted and superconformal indices}
\label{sec:6}

Let us now consider the near-horizon of spacetime M2-branes wrapped on two-dimensional compact toric surfaces $\Sigma$, corresponding holographically to Chern--Simons-matter theories on the background $S^1 \times \Sigma$. The details of this background were summarized in \cite[Section~6]{Cassia:2025aus}; we begin with a brief recap below.

\subsubsection*{Summary of wrapped M2-branes}

In the wrapped brane case, the near-horizon geometry becomes
\be
	M_{11} = \mathrm{AdS}_2 \times M\ ,
\ee
where $M$ is the total space of a fibration of $L$ over the surface $\Sigma$, which we assume to be of the form $\Sigma=\WPL$, i.e.\ a weighted projective line (or \emph{spindle}). Taking the cone of each fiber $L$, gives rise to another fibration whose total space we denote as $Y$. Both $M$ and $Y$ can then be described via the diagrams
\be
\begin{array}{ccc}
 L & \to & M \\
 && \downarrow \\
 && \Sigma
\end{array}\ ,
\qquad \qquad \qquad
\begin{array}{ccc}
 C(L) & \to & Y \\
 && \downarrow \\
 && \Sigma
\end{array}\ .
\ee
In what follows, we will use indices $i$ for the coordinates in the fiber, and indices $+,-$ for the homogeneous coordinates in the base. The collective coordinates on the total space of the fibration will be denoted instead as $I=\{+,-,i\}$. We also recall that the spindle can be described as the toric quotient $\BC^2//\BC^\times$ with charges $a,b\in\BZ_{>0}$ for the $\BC^\times$-action on the two coordinates. In this case, we can assume that $Y$ also admits a description as a toric quotient of the form $Y=(\BC^2\times C(L))//\BC^\times$ where the $\BC^\times$ acts on the coordinates of $C(L)$ with charges $Q_i$, and on $\BC^2$ as before. We can then describe the fibration via the charge matrix
\be
\label{eq:fibratioYchargesQ}
	Q = (a, b, - n_1,-n_2, \dots )\ , \qquad \sum_i n_i = a + b\ ,
\ee
where we take $n_i$ to be positive integers whose sum is constrained to be $a+b$ by the CY condition on $Y$.

In the effective 4d supergravity, after compactification over $L$, the background can be seen as the near-horizon of asymptotically AdS$_4$ black holes, admitting a general set of electric and magnetic charges responcible for parametrizing the fibration of $L$ over $\Sigma$, see \cite{Cacciatori:2009iz,DallAgata:2010ejj,Hristov:2010ri,Halmagyi:2013sla, Hristov:2018spe, Hristov:2019mqp, Ferrero:2020twa, Couzens:2021cpk, Ferrero:2021etw, Hristov:2023rel}. The dual field theory interpretation is of an RG flow from a theory on $S^1 \times \Sigma$ in the UV to a 1d superconformal quantum mechanics in the IR. Consequently, the dual 3d partition function in this case has the interpretation of an index: when $\Sigma$ is the two-sphere the twisted or superconformal index, generalized by the twisted and anti-twisted spindle indices when $\Sigma = \WPL$, see \cite{Kim:2009wb,Kapustin:2011jm,Beem:2012mb,Benini:2015noa,Benini:2015eyy,Hosseini:2016tor,Hosseini:2016ume,Hosseini:2022vho,Bobev:2022eus,Bobev:2024mqw,Inglese:2023wky,Colombo:2024mts}.

As discussed in more detail in \cite{Couzens:2018wnk,Gauntlett:2018dpc,Hosseini:2019use,Hosseini:2019ddy,Gauntlett:2019roi}, large $N$ holography dictates the condition 
\be
\label{eq;holorelspindle}
	\frac1{\pi^2}\, \left. \frac{\partial N^Y_{0, 0} (\lambda, \epsilon)}{\partial \lambda^I} \right|_{\lam^i = \lam^i_\text{mes.}} = \frac{Q_I}{a b}\, N\ ,
\ee
where we take derivatives w.r.t.\ all fiber and base K\"ahler parameters $\lam^I$ but we only impose the mesonic twist condition in the fiber, which means that we keep the size of the base finite.~\footnote{Again, we do not allow for mixing with baryonic symmetries in this condition, see \cite{Hosseini:2025mgf} for relaxing this assumption.}
Moreover, we denote as $Q_I$ the charge of the $I$-th homogeneous coordinate as defined in \eqref{eq:fibratioYchargesQ}.

We refer the reader to \cite[Section~5]{Cassia:2025aus} for more details on the evaluation of the equivariant volume and resulting topological string constant maps terms, for this geometry. Here we reproduce the main results, using the identity \eqref{eq:mesonictwistvolume},
\be
	\BV_Y (\lam, \NN, \e ) = \frac1{\NN} \left( \BV_X (\lam^i_{+, \text{mes.}}, \e_i^+) -  \BV_X (\lam^i_{-, \text{mes.}}, \e_i^-) \right)\ ,
\ee
with
\be
\label{eq:shiftedepsilons}
 \e_{i}^+ := \e_i+\frac{n_i}{a}\e_+\,, \qquad \e_{i}^- := \e_i+\frac{n_i}{b}\e_-\ ,
\ee
and
\be
\label{eq:shiftedlambdas}
 \lam^i_{+, \text{mes.}} := \lam^i_\text{mes.} + \frac{\NN}{a\, n\, \e_i^+}\, \lam^-\ ,
 \qquad
 \lam^i_{-, \text{mes.}} := \lam^i_\text{mes.} - \frac{\NN}{b\, n\, \e_i^-}\, \lam^+\ ,
\ee
where the additional parameters $\lam^\pm$, $\e_\pm$ and $\NN:=a\e_--b\e_+$ correspond to the parameters in the base, while $n$ is the dimension of the (possibly ineffective) toric action on $X$ (not to be confused with the fluxes $n_i$ themselves). These expressions lead to the genus-zero constant maps term
\be
\label{eq:mesonicgenuszeroafterparametrizationfibration}
\begin{aligned}
 N^Y_{0, 0} &=  \frac1{\NN} \left( N_{0, 0} ^{X, \text{mes.}} (\lam^i_{+, \text{mes.}}, \e_i^+) -  N_{0, 0} ^{X, \text{mes.}} (\lam^i_{-,  \text{mes.}}, \e_i^-) \right) \\
 & = \frac{1}{6\, \NN}\, \left( C_X (\e^+_i)\, \Big(\sum_i\e_i^+\lam^i_{+, \text{mes.}}\Big)^3 -  C_X (\e^-_i)\, \Big(\sum_i\e_i^-\lam^i_{-, \text{mes.}}\Big)^3 \right)\ .
\end{aligned}
\ee

\subsection{Leading order in \texorpdfstring{$N$}{N}}
Let us now follow the logic of Section~\ref{sec:2} and infer the relation between the topological string and the dual index at leading order, based on the known supergravity and field theory results. See in particular the supergravity gluing constructions in \cite{Hosseini:2019iad,Hosseini:2021fge,Faedo:2021nub,Boido:2022mbe,Hristov:2023rel}. Most of the following formulas and results have been discussed in greater detail in \cite{Martelli:2023oqk} (with some differing conventions and notations), so we keep the discussion short.

In this case, in order to match with the holographic computations in \cite{Martelli:2023oqk}, we impose the identification of equivariant parameters
\be
	-2 b\, \e_+ = 2 a\, \e_- = \NN\ ,
\ee
which we can use to express $\e_\pm$ in terms of $\NN$.
Then we can infer from the effective four-dimensional supergravity the supersymmetric constraints,
\be
\label{eq:susyconstroncharges}
    \begin{array}{ccrl}
    \sum_i n_i = a + b\ , & \hspace{30pt} & \sum_i \e_i &= 2 -(\e_++\e_-) \\
     &&&= 2+\frac{\NN}{2b}-\frac{\NN}{2a} \\
     &&&= 2 + \frac{a - b}{2 a b}\, \NN\ ,
    \end{array}
\ee
which can be rewritten as
\be
	\sum_i \e_i^+ = 2 - \frac{\NN}{a}\ , \qquad \sum_i \e_i^- = 2 + \frac{\NN}{b}\ ,
\ee
using the shifted equivariant parameters, \eqref{eq:shiftedepsilons}.

Additionally, since we are using the mesonic condition on the fiber $X=C(L)$, we can use the same parametrization for the $\lam^i$ variables as in \eqref{eq:univparametrization},
\be
	\lam^i_{\text{mes.}} = \tilde \mu\ , \qquad \forall i\ ,
\ee
while keeping the spindle parameters $\lam^+$ and $\lam^-$ independent.
At this stage it might seem that the constant maps term, \eqref{eq:mesonicgenuszeroafterparametrizationfibration}, depends on three independent parameters, $\tilde \mu, \lam_\pm$, but this is not the case. The dependence on $\lam$ comes through the summation $\sum_i\e^\pm_i\lam^i_{\pm,\text{mes.}}$ with the equivariant parameters, which we can write explicitly as
\be
\label{eq:simplifspindlelam}
\begin{aligned}
	\sum_i\e_i^+\lam^i_{+, \text{mes.}}
    &= \Big(2 - \frac{\NN}{a}\Big)\, \tilde \mu + \frac{\NN}{a}\, \lam^- =:  \tilde \mu_+\ , \\
	\sum_i\e_i^-\lam^i_{-, \text{mes.}}
    &= \Big(2 + \frac{\NN}{b}\Big)\, \tilde \mu - \frac{\NN}{b}\, \lam^+ =:  \tilde \mu_-\ ,
\end{aligned}
\ee
so that only two linear combinations of the three parameters appear.
We define these two independent parameters as $\tilde{\mu}_\pm$ for short.~\footnote{Note that we could have simply set $\tilde\mu = 0$ from the outset in agreement with \cite{Martelli:2023oqk}, but we show that this is not necessary in order to arrive at the same final result. In addition, we flipped the sign between $\lam_+$ and $\tilde \mu_-$ for a purely conventional purpose, since the sign ambiguity in the consequent extremization forces us to make a choice for the chamber of $\tilde\mu_\pm$.}

We can thus rewrite the genus-zero constant maps term, \eqref{eq:mesonicgenuszeroafterparametrizationfibration}, as
\be
	N_{0,0}^Y (\tilde \mu_\pm, \e_i^\pm) =\frac{1}{6 \NN}\, \left( C_X (\e_i^+)\, \tilde \mu_+^3 -  C_X (\e_i^-)\, \tilde \mu_-^3 \right)\ .
\ee
Following the logic of Section~\ref{sec:2}, we can rewrite the large $N$ holographic relation for the spindle fibration, \eqref{eq;holorelspindle}, as a saddle point of the integral transform,
\be
\label{eq:finalintegrationindices}
\begin{aligned}
	Z_M^\text{pert} (\Delta, \NN, N_{\rm M2}) &= \int \mathd \tilde \mu_+ \mathd  \tilde \mu_-\, \exp  \left( \frac1{\pi^2}\, N_{0,0}^Y (\tilde \mu_\pm, \Delta^\pm)  -  \frac{N}{a b}\, (-\sum_in_i \lam^i_\text{mes.} + a \lambda^+ + b \lambda^- ) \right) \\
	&=  \int \mathd \tilde \mu_+ \mathd \tilde \mu_-\, \exp  \left( \frac1{6 \pi^2\, \NN}\, \left( C_X (\Delta_i^+)\, \tilde \mu_+^3 -  C_X (\Delta_i^-)\, \tilde \mu_-^3 \right)  - \frac{N}{\NN} (\tilde \mu_+ - \tilde \mu_-) \right)\ ,
\end{aligned}
\ee
where the second line follows from \eqref{eq:simplifspindlelam}, and we again identified $\e_i$ with $\Delta_i$ as in \eqref{eq:mesonictwistimplication}. From the previous supersymmetric identities \eqref{eq:susyconstroncharges} and \eqref{eq:shiftedepsilons}, we therefore have
\be
\label{eq:bhsusyconstrft}
	\Delta_{i}^{\pm} = \Delta_i \mp \frac{n_i}{2 a b}\, \NN\ ,
    \hspace{30pt}
    \sum_i \Delta_i = 2 + \frac{a - b}{2 a b}\, \NN\ .
\ee
In order to be able to match to both the twisted index and superconformal index computations, we perform analytic continuation in the charge $b$, which up to now was assumed to be a positive integer. We will allow $b$ to take negative values as well, and we define a new parameter $\sigma$ as the sign of $b$, namely
\be
    \sigma := \frac{b}{|b|}.
\ee
We note that, in these convention, $\sigma = 1$ corresponds to the so called \emph{twist} condition, while $\sigma = -1$ to the \emph{anti-twist}, \cite{Ferrero:2021etw}, generalizing to the spindle case the twisted and superconformal indices (at $a = |b| = 1$), respectively.

Finally, we can look at the leading order at large $N$,~\footnote{Here we pick the chamber $\tilde \mu_+ > 0$, $ \sigma \tilde\mu_- > 0$, which matches the supergravity result and agrees with the analogous choice in \cite{Martelli:2023oqk}.}
\be
	- \log Z_M^\text{pert} (\Delta, \NN, N_{\rm M2}) \stackrel{\text{saddle pt.}}{\approx}  \cF_M(\Delta, \NN, N) = \frac{2 \sqrt{2} \pi\, N^{3/2}}{3\,  \NN} \left(\frac{1}{\sqrt{C_X (\Delta^+)}} - \frac{\sigma}{\sqrt{C_X (\Delta^-)}}  \right)\ .
\ee
This is in agreement with field theory and supergravity calculations, \cite{Hosseini:2019iad,Faedo:2021nub,Hosseini:2022vho,Boido:2022mbe,Colombo:2024mts}, which can be now written as
\be
\label{eq:leadingorderindexfromtopostring}
	\cF_M(\Delta, \NN, N) = \frac1{2\, \NN} \left(\cF_L(\Delta^+, N) - \sigma\, \cF_L(\Delta^-, N)  \right)\ .
\ee
We have thus reproduced the main result in \cite{Martelli:2023oqk} in our notation, which establishes geometrically the gravitational block decomposition in \cite{Hosseini:2019iad,Faedo:2021nub,Boido:2022mbe}.

\subsection{Airy function structure}

Ignoring the subleading corrections in the exact wrapped-brane charges and the topological string partition function, we could still go beyond the large $N$ limit in field theory by simply performing the exact integration in \eqref{eq:finalintegrationindices} instead of the saddle-point approximation. We again assume the integration contour related to the integral representation of the Airy functions of first and second kind, c.f.\ \eqref{eq:airyasympt}, and ignoring a constant prefactor (in $N$) arrive at
\be
\begin{aligned}
	\text{twist}: \quad Z_M^\text{pert} &  \simeq {\rm Ai} \Big[ \left( \frac{\NN^2\, C_{X} (\Delta_+)}{2 \pi^2}\right)^{-1/3}\, N \Big] \times {\rm Bi} \Big[ \left( \frac{\NN^2\, C_{X} (\Delta_-)}{2 \pi^2}\right)^{-1/3}\, N \Big]\ , \\
	\text{anti-twist}: \quad Z_M^\text{pert} &  \simeq {\rm Ai} \Big[ \left( \frac{\NN^2\, C_{X} (\Delta_+)}{2 \pi^2}\right)^{-1/3}\, N \Big] \times {\rm Ai} \Big[ \left( \frac{\NN^2\, C_{X} (\Delta_-)}{2 \pi^2}\right)^{-1/3}\, N \Big]\ .
\end{aligned}
\ee
Up to a shift in the value of $N$, which comes from the subleading corrections we ignored, this is in exact agreement with the result in \cite{Hristov:2022lcw} based on HD supergravity building blocks.~\footnote{The precise match uses the identification $\NN^\text{here} = 2\, \omega^\text{there}$. This is also in agreement with the available explicit localization results in different limiting cases, see \cite{Hosseini:2022vho,Bobev:2023lkx,Bobev:2024mqw}.} Note that the Airy function of second kind, ${\rm Bi}$, has a slightly different asymptotic expansion with a leading exponent of the opposite sign with respect to \eqref{eq:airyasympt}, see again \cite{Hristov:2022lcw}. This sign is precisely related to the overall negative sign in the second term of \eqref{eq:leadingorderindexfromtopostring} in the twist case, i.e.\ $\sigma = 1$.

We do not attempt here to further rederive the exact answer in \cite{Hristov:2022lcw}, noting that our present geometric construction nevertheless strongly supports it. In order to fully explore the subleading corrections, we need to calculate the exact wrapped-brane charges incorporating eight-derivative corrections in 11d, as well as revisit the genus-one refinement we introduced in Section~\ref{sec:3}. We aim to address this problem in the near future.

\section{D3-brane models}
\label{sec:7}
In this section, we sketch the proposed generalization of the equivariant topological string formalism applied to other brane models, such as D-branes in type IIB and massive type IIA string theories. The general idea was already mentioned in \cite[Section~8]{Cassia:2025aus}, and here we aim to present a simple explicit calculation that supports it. We are going to focus specifically on D3-branes, where the relation between $X$ and $L$ is exactly the same as for M2-branes.

\subsection{Spacetime filling branes}
The D3-brane description closely resembles the M2-brane discussion in the rest of this paper, due to the fact that the near-horizon geometry of multiple D3-branes also results in a compact Sasakian manifold $L$, this time five- instead of seven-dimensional.
Because of this, we can consider manifolds $X$ which are toric Calabi--Yau three-folds obtained as resolutions of cones over a 5-dimensional Sasakian base $L$.
Since the equivariant framework can be generalized to arbitrary dimensions, essentially all geometric calculations for the M2-brane models have a direct analog for D3-branes, as already explored at some length in the references \cite{Martelli:2005tp,Butti:2005vn,Butti:2005ps,Martelli:2006yb,Amariti:2011uw,Couzens:2018wnk,Gauntlett:2018dpc,Hosseini:2019use,Hosseini:2019ddy,Gauntlett:2019roi,Gauntlett:2019pqg,Boido:2022mbe,Martelli:2023oqk,Colombo:2023fhu}. In turn, many of the calculations and explicit examples in \cite{Cassia:2025aus} are immediately applicable to D3-brane systems, including examples of fibrations over a surface $\Sigma$.

Importantly, D3-branes are fundamental objects in type IIB string theory, which according to our general proposal, correspond to topological strings expanded at degree two in $\lam$, rather than three. We thus arrive at
\be
\label{eq:D3branepf}
    Z_L = \int \mathd \lam\, \exp \left( F^\text{top,pert}_{X,({\rm deg} = 2)} (\lam, \e; g_s) - \lam\, N_{\rm D3}  \right)\ , 
\ee
where, following \cite[Section~2]{Cassia:2025aus}, we can define the leading order equivariant constant maps term of second degree,
\be
\label{eq:seconddegreetopostring}
	F^\text{top,pert}_{X,(2)} (\lam, \e; g_s) := \frac1{2 g_s^2}\, \sum_{i, j} \frac{\partial^2 \BV (\lam, \e)}{\partial \lam^i \partial \lam^j} \Big|_{\lam = 0}\, \lam^i \lam^j\ ,
\ee
where we ignore constant and non-perturbative terms in $\lam$. Again, $X$ (now a three-fold) can be described as a resolution of the cone over $L$, and the mesonic twist condition corresponds to the blow-down to the conical space, $C(L)$. Note that we are not specifying here in detail the meaning of the partition function $Z_L$, which due to the specifics of 4d superconformal field theories and the background space, might be ambiguously defined, see \cite{Papadimitriou:2017kzw}. At present we are simply pursuing the difference between the saddle-point evaluation and the full integral, in order to show more directly what our proposal amounts to.

Given the similarities to Section~\ref{sec:2}, and the fact that holographic relations and R-charge assignments in the dual four-dimensional theory obey the same constraint,
\be
	\sum_i \Delta_i = 2\ ,
\ee
we are going to assume once again the holographic relation
\be
	\e_i = \Delta_i\ ,
\ee
upon imposing the mesonic twist and the constraint \eqref{eq:mesonicconstraint}.
Furthermore, still using the mesonic twist, it is straightforward to derive an explicit expression for the leading order term in this case,
\be
    F^\text{top,pert}_{X,(2)} (\lam_\text{mes.}, \e; g_s) =
    \frac1{2 g_s^2}\,(\e_i \lam_\text{mes.}^i)^2\, C_X (\e)\ , 
\ee
such that with the choice of the universal parametrization
\be
	\lam_\text{mes.}^i = \tilde \mu\ , \qquad  \forall i\ ,
\ee
we obtain a further simplification
\be
	F^\text{top,pert}_{X,(2)} (\lam_\text{mes.}, \e; g_s)
    = \frac{2\, C_X (\e)}{g_s^2}\, \tilde \mu^2\ .
\ee
At leading order, by definition we use $N_{\rm D3} = N$, so that in turn we can evaluate \eqref{eq:D3branepf} as
\be
	Z_L^\text{pert} (\Delta, N) = \int \mathd \tilde \mu\, \exp \left(\frac{2\, C_X (\Delta)}{g_s^2}\, \tilde \mu^2 -2\, \tilde \mu\, N  \right)\ , 
\ee
such that
\be
-\log Z_L^\text{pert} \simeq \frac{g_s^2\, N^2}{2\, C_X(\Delta)} \ ,
\ee
up to constant in $N$, which we have already ignored in \eqref{eq:seconddegreetopostring}. Due to the fact that we have a Gaussian integral, we found that the saddle-point evaluation and the exact integration precisely agree, as far as the $N$-dependence is concerned. The result can be successfully matched with $a$-maximization in field theory, \cite{Intriligator:2003jj}, as shown in many of the references cited above.

It is straightforward to see that the analysis of the wrapped D3-branes over toric surfaces $\Sigma$ again follows closely the steps from the previous section, see \cite{Martelli:2023oqk}. Once more, the appearance of Gaussian integral means that we recover the saddle-point results regarding the $N$ dependence. The additional predictions of our present approach, when applied to D3-branes, can thus be only discerned from the previously known results when we include the analysis of constant prefactors, which we hope to explore elsewhere.

\section{Discussion and outlook}
\label{sec:conclusion}
We conclude with a selected list of topics that extend the relation between the topological string and field theory or supergravity proposed here, and which we hope to revisit in the future.

\subsubsection*{Higher-genus constant maps}
In the main prediction we derived for the dual field theory partition function, see \eqref{eq:centralresult}, we neglected constant prefactors in $N$, as well as non-perturbative corrections. While the latter require a more detailed understanding of equivariant Gromov--Witten invariants (see the discussion in \cite{Cassia:2025aus}), the former are presently much more accessible. They arise from the contributions of constant maps at higher genus, $\mathfrak{g} > 1$. These terms, in the absence of refinement, were written down in \cite{Cassia:2025aus} and are straightforward to evaluate using the equivariant volume $\BV_X$.

It is, in fact, quite interesting to note their resemblance to the known corrections to the Airy function, typically denoted by $A(\Delta)$; see \cite{Hanada:2012si,Moriyama:2014gxa,Nosaka:2015iiw}. However, in order to achieve an exact match between field theory and topological strings, it is necessary to fully understand the refined contributions that can appear at the same order. As indicated in \eqref{eq:topostringrefinement}, there are $\mathfrak{g}$ distinct such terms at each genus. These contributions are more suitably analyzed via the effective supergravity perspective, where they are expected to originate from a calculation generalizing \cite{Dedushenko:2014nya}.

\subsubsection*{TS/ST correspondence}
Given the current predictions for the squashed sphere partition function in M2-brane theories, it is natural to explore their connection with the topological string/spectral theory (TS/ST) correspondence, as introduced in \cite{Grassi:2014zfa,Marino:2015nla,Codesido:2015dia} and subsequent works. While the topological string framework employed in those references shares the same formal genus expansion, it does not incorporate the equivariant refinement. As a result, the constant map contributions on non-compact toric Calabi–Yau manifolds remain divergent and cannot be directly compared with the present proposal. Nevertheless, it appears highly plausible that the equivariant extension introduced in \cite{Cassia:2025aus} can be integrated with the TS/ST correspondence. This would not only extend the latter framework but also establish a more direct connection with the approach developed here.

\subsubsection*{Towards proof of holography}
In this work, we have explored the interplay between topological string theory, supergravity, and their dual field theory counterparts. When combined with the spectral theory framework mentioned above, these connections form an intricate network of dualities and equivalent descriptions of the same underlying physics. This growing web of correspondences strengthens the prospect—at least within the BPS sector—of formulating a complete quantum gravity, or more precisely, quantum geometry, computation from first principles, independently of the holographic field theory description. Such a formulation would open the door to pursuing a formal proof of the AdS/CFT correspondence for BPS observables. We believe that the present results represent a conceptual step toward this ambitious goal.

\subsection*{Acknowledgements}
We would like to thank Fridrik Freyr Gautason, Seyed Morteza Hosseini, Naotaka Kubo, Yi Pang, Jesse van Muiden, and Alberto Zaffaroni for useful discussions. We gratefully acknowledge support during the MATRIX Program “New Deformations of Quantum Field and Gravity Theories,” (Creswick, 22 Jan – 2 Feb 2024), where this project was initiated.
The work of L.C.\ was supported by the ARC Discovery Grant DP210103081.
The study of K.H.\ is financed by the European Union- NextGenerationEU, through the National Recovery and Resilience Plan of the Republic of Bulgaria, project No BG-RRP-2.004-0008-C01.

%%%%%%%%%%%%%%%%%%%%%%%%%%%%%%%%%%%%%
\bibliographystyle{JHEP}
\bibliography{refs.bib}

\providecommand{\href}[2]{#2}\begingroup\raggedright\begin{thebibliography}{100}

\bibitem{Cassia:2025aus}
L.~Cassia and K.~Hristov, {\it {Constant maps in equivariant topological
  strings and geometric modeling of fluxes}},  {\em J. Phys. A} {\bf 58}
  (2025), no.~49 495201, [\href{http://arxiv.org/abs/2502.20444}{{\tt
  arXiv:2502.20444}}].

\bibitem{Martelli:2023oqk}
D.~Martelli and A.~Zaffaroni, {\it {Equivariant localization and holography}},
  {\em Lett. Math. Phys.} {\bf 114} (2024), no.~1 15,
  [\href{http://arxiv.org/abs/2306.03891}{{\tt arXiv:2306.03891}}].

\bibitem{Colombo:2023fhu}
E.~Colombo, F.~Faedo, D.~Martelli, and A.~Zaffaroni, {\it {Equivariant volume
  extremization and holography}},  {\em JHEP} {\bf 01} (2024) 095,
  [\href{http://arxiv.org/abs/2309.04425}{{\tt arXiv:2309.04425}}].

\bibitem{Gautason:2025plx}
F.~F. Gautason and J.~van Muiden, {\it {Ensembles in M-theory and holography}},
   {\em JHEP} {\bf 11} (2025) 078, [\href{http://arxiv.org/abs/2505.21633}{{\tt
  arXiv:2505.21633}}].

\bibitem{Aharony:2008ug}
O.~Aharony, O.~Bergman, D.~L. Jafferis, and J.~Maldacena, {\it {N=6
  superconformal Chern-Simons-matter theories, M2-branes and their gravity
  duals}},  {\em JHEP} {\bf 10} (2008) 091,
  [\href{http://arxiv.org/abs/0806.1218}{{\tt arXiv:0806.1218}}].

\bibitem{Benini:2009qs}
F.~Benini, C.~Closset, and S.~Cremonesi, {\it {Chiral flavors and M2-branes at
  toric CY4 singularities}},  {\em JHEP} {\bf 02} (2010) 036,
  [\href{http://arxiv.org/abs/0911.4127}{{\tt arXiv:0911.4127}}].

\bibitem{Cremonesi:2010ae}
S.~Cremonesi, {\it {Type IIB construction of flavoured ABJ(M) and fractional M2
  branes}},  {\em JHEP} {\bf 01} (2011) 076,
  [\href{http://arxiv.org/abs/1007.4562}{{\tt arXiv:1007.4562}}].

\bibitem{Kapustin:2009kz}
A.~Kapustin, B.~Willett, and I.~Yaakov, {\it {Exact Results for Wilson Loops in
  Superconformal Chern-Simons Theories with Matter}},  {\em JHEP} {\bf 03}
  (2010) 089, [\href{http://arxiv.org/abs/0909.4559}{{\tt arXiv:0909.4559}}].

\bibitem{Drukker:2010nc}
N.~Drukker, M.~Marino, and P.~Putrov, {\it {From weak to strong coupling in
  ABJM theory}},  {\em Commun. Math. Phys.} {\bf 306} (2011) 511--563,
  [\href{http://arxiv.org/abs/1007.3837}{{\tt arXiv:1007.3837}}].

\bibitem{Herzog:2010hf}
C.~P. Herzog, I.~R. Klebanov, S.~S. Pufu, and T.~Tesileanu, {\it {Multi-Matrix
  Models and Tri-Sasaki Einstein Spaces}},  {\em Phys. Rev. D} {\bf 83} (2011)
  046001, [\href{http://arxiv.org/abs/1011.5487}{{\tt arXiv:1011.5487}}].

\bibitem{Santamaria:2010dm}
R.~C. Santamaria, M.~Marino, and P.~Putrov, {\it {Unquenched flavor and
  tropical geometry in strongly coupled Chern-Simons-matter theories}},  {\em
  JHEP} {\bf 10} (2011) 139, [\href{http://arxiv.org/abs/1011.6281}{{\tt
  arXiv:1011.6281}}].

\bibitem{Jafferis:2010un}
D.~L. Jafferis, {\it {The Exact Superconformal R-Symmetry Extremizes Z}},  {\em
  JHEP} {\bf 05} (2012) 159, [\href{http://arxiv.org/abs/1012.3210}{{\tt
  arXiv:1012.3210}}].

\bibitem{Hama:2010av}
N.~Hama, K.~Hosomichi, and S.~Lee, {\it {Notes on SUSY Gauge Theories on
  Three-Sphere}},  {\em JHEP} {\bf 03} (2011) 127,
  [\href{http://arxiv.org/abs/1012.3512}{{\tt arXiv:1012.3512}}].

\bibitem{Hama:2011ea}
N.~Hama, K.~Hosomichi, and S.~Lee, {\it {SUSY Gauge Theories on Squashed
  Three-Spheres}},  {\em JHEP} {\bf 05} (2011) 014,
  [\href{http://arxiv.org/abs/1102.4716}{{\tt arXiv:1102.4716}}].

\bibitem{Martelli:2011qj}
D.~Martelli and J.~Sparks, {\it {The large N limit of quiver matrix models and
  Sasaki-Einstein manifolds}},  {\em Phys. Rev. D} {\bf 84} (2011) 046008,
  [\href{http://arxiv.org/abs/1102.5289}{{\tt arXiv:1102.5289}}].

\bibitem{Cheon:2011vi}
S.~Cheon, H.~Kim, and N.~Kim, {\it {Calculating the partition function of N=2
  Gauge theories on $S^3$ and AdS/CFT correspondence}},  {\em JHEP} {\bf 05}
  (2011) 134, [\href{http://arxiv.org/abs/1102.5565}{{\tt arXiv:1102.5565}}].

\bibitem{Jafferis:2011zi}
D.~L. Jafferis, I.~R. Klebanov, S.~S. Pufu, and B.~R. Safdi, {\it {Towards the
  F-Theorem: N=2 Field Theories on the Three-Sphere}},  {\em JHEP} {\bf 06}
  (2011) 102, [\href{http://arxiv.org/abs/1103.1181}{{\tt arXiv:1103.1181}}].

\bibitem{Fuji:2011km}
H.~Fuji, S.~Hirano, and S.~Moriyama, {\it {Summing Up All Genus Free Energy of
  ABJM Matrix Model}},  {\em JHEP} {\bf 08} (2011) 001,
  [\href{http://arxiv.org/abs/1106.4631}{{\tt arXiv:1106.4631}}].

\bibitem{Imamura:2011wg}
Y.~Imamura and D.~Yokoyama, {\it {N=2 supersymmetric theories on squashed
  three-sphere}},  {\em Phys. Rev. D} {\bf 85} (2012) 025015,
  [\href{http://arxiv.org/abs/1109.4734}{{\tt arXiv:1109.4734}}].

\bibitem{Marino:2011eh}
M.~Marino and P.~Putrov, {\it {ABJM theory as a Fermi gas}},  {\em J. Stat.
  Mech.} {\bf 1203} (2012) P03001, [\href{http://arxiv.org/abs/1110.4066}{{\tt
  arXiv:1110.4066}}].

\bibitem{Martelli:2011fu}
D.~Martelli, A.~Passias, and J.~Sparks, {\it {The gravity dual of
  supersymmetric gauge theories on a squashed three-sphere}},  {\em Nucl. Phys.
  B} {\bf 864} (2012) 840--868, [\href{http://arxiv.org/abs/1110.6400}{{\tt
  arXiv:1110.6400}}].

\bibitem{Hatsuda:2012dt}
Y.~Hatsuda, S.~Moriyama, and K.~Okuyama, {\it {Instanton Effects in ABJM Theory
  from Fermi Gas Approach}},  {\em JHEP} {\bf 01} (2013) 158,
  [\href{http://arxiv.org/abs/1211.1251}{{\tt arXiv:1211.1251}}].

\bibitem{Hatsuda:2013oxa}
Y.~Hatsuda, M.~Marino, S.~Moriyama, and K.~Okuyama, {\it {Non-perturbative
  effects and the refined topological string}},  {\em JHEP} {\bf 09} (2014)
  168, [\href{http://arxiv.org/abs/1306.1734}{{\tt arXiv:1306.1734}}].

\bibitem{Moriyama:2014gxa}
S.~Moriyama and T.~Nosaka, {\it {Partition Functions of Superconformal
  Chern-Simons Theories from Fermi Gas Approach}},  {\em JHEP} {\bf 11} (2014)
  164, [\href{http://arxiv.org/abs/1407.4268}{{\tt arXiv:1407.4268}}].

\bibitem{Nosaka:2015iiw}
T.~Nosaka, {\it {Instanton effects in ABJM theory with general R-charge
  assignments}},  {\em JHEP} {\bf 03} (2016) 059,
  [\href{http://arxiv.org/abs/1512.02862}{{\tt arXiv:1512.02862}}].

\bibitem{Hatsuda:2016uqa}
Y.~Hatsuda, {\it {ABJM on ellipsoid and topological strings}},  {\em JHEP} {\bf
  07} (2016) 026, [\href{http://arxiv.org/abs/1601.02728}{{\tt
  arXiv:1601.02728}}].

\bibitem{Kubo:2019ejc}
N.~Kubo and S.~Moriyama, {\it {Hanany-Witten Transition in Quantum Curves}},
  {\em JHEP} {\bf 12} (2019) 101, [\href{http://arxiv.org/abs/1907.04971}{{\tt
  arXiv:1907.04971}}].

\bibitem{Hosseini:2019and}
S.~M. Hosseini, C.~Toldo, and I.~Yaakov, {\it {Supersymmetric R{\'e}nyi entropy
  and charged hyperbolic black holes}},  {\em JHEP} {\bf 07} (2020) 131,
  [\href{http://arxiv.org/abs/1912.04868}{{\tt arXiv:1912.04868}}].

\bibitem{Kubo:2020qed}
N.~Kubo, {\it {Fermi gas approach to general rank theories and quantum
  curves}},  {\em JHEP} {\bf 10} (2020) 158,
  [\href{http://arxiv.org/abs/2007.08602}{{\tt arXiv:2007.08602}}].

\bibitem{Chester:2021gdw}
S.~M. Chester, R.~R. Kalloor, and A.~Sharon, {\it {Squashing, Mass, and
  Holography for 3d Sphere Free Energy}},  {\em JHEP} {\bf 04} (2021) 244,
  [\href{http://arxiv.org/abs/2102.05643}{{\tt arXiv:2102.05643}}].

\bibitem{Bobev:2022jte}
N.~Bobev, J.~Hong, and V.~Reys, {\it {Large N Partition Functions, Holography,
  and Black Holes}},  {\em Phys. Rev. Lett.} {\bf 129} (2022), no.~4 041602,
  [\href{http://arxiv.org/abs/2203.14981}{{\tt arXiv:2203.14981}}].

\bibitem{Bobev:2022eus}
N.~Bobev, J.~Hong, and V.~Reys, {\it {Large N partition functions of the ABJM
  theory}},  {\em JHEP} {\bf 02} (2023) 020,
  [\href{http://arxiv.org/abs/2210.09318}{{\tt arXiv:2210.09318}}].

\bibitem{Geukens:2024zmt}
S.~Geukens and J.~Hong, {\it {Subleading analysis for S$^{3}$ partition
  functions of $ \mathcal{N} $ = 2 holographic SCFTs}},  {\em JHEP} {\bf 06}
  (2024) 190, [\href{http://arxiv.org/abs/2405.00845}{{\tt arXiv:2405.00845}}].

\bibitem{Kubo:2024qhq}
N.~Kubo, T.~Nosaka, and Y.~Pang, {\it {Exact large N expansion of mass deformed
  ABJM theory on squashed sphere}},  {\em JHEP} {\bf 02} (2025) 106,
  [\href{http://arxiv.org/abs/2411.07334}{{\tt arXiv:2411.07334}}].

\bibitem{Kubo:2025jxi}
N.~Kubo, {\it {Five-brane webs, 3d $ \mathcal{N} $ = 2 theories and quantum
  curves}},  {\em JHEP} {\bf 05} (2025) 103,
  [\href{http://arxiv.org/abs/2501.04146}{{\tt arXiv:2501.04146}}].

\bibitem{Bobev:2025ltz}
N.~Bobev, P.-J. De~Smet, J.~Hong, V.~Reys, and X.~Zhang, {\it {An Airy tale at
  large N}},  {\em JHEP} {\bf 07} (2025) 123,
  [\href{http://arxiv.org/abs/2502.04606}{{\tt arXiv:2502.04606}}].

\bibitem{Kubo:2025dot}
N.~Kubo, T.~Nosaka, and Y.~Pang, {\it {Exact large N expansion of N=4 circular
  quiver Chern-Simons theories}},  {\em Phys. Rev. D} {\bf 112} (2025), no.~4
  046023, [\href{http://arxiv.org/abs/2504.04402}{{\tt arXiv:2504.04402}}].

\bibitem{Bobev:2020egg}
N.~Bobev, A.~M. Charles, K.~Hristov, and V.~Reys, {\it {The Unreasonable
  Effectiveness of Higher-Derivative Supergravity in AdS$_4$ Holography}},
  {\em Phys. Rev. Lett.} {\bf 125} (2020), no.~13 131601,
  [\href{http://arxiv.org/abs/2006.09390}{{\tt arXiv:2006.09390}}].

\bibitem{Bobev:2021oku}
N.~Bobev, A.~M. Charles, K.~Hristov, and V.~Reys, {\it {Higher-derivative
  supergravity, AdS$_{4}$ holography, and black holes}},  {\em JHEP} {\bf 08}
  (2021) 173, [\href{http://arxiv.org/abs/2106.04581}{{\tt arXiv:2106.04581}}].

\bibitem{Hristov:2021qsw}
K.~Hristov, {\it {4d $ \mathcal{N} $ = 2 supergravity observables from
  Nekrasov-like partition functions}},  {\em JHEP} {\bf 02} (2022) 079,
  [\href{http://arxiv.org/abs/2111.06903}{{\tt arXiv:2111.06903}}].

\bibitem{Hristov:2022lcw}
K.~Hristov, {\it {ABJM at finite N via 4d supergravity}},  {\em JHEP} {\bf 10}
  (2022) 190, [\href{http://arxiv.org/abs/2204.02992}{{\tt arXiv:2204.02992}}].

\bibitem{Hristov:2022plc}
K.~Hristov, {\it {Maximally symmetric nuts in 4d \ensuremath{\mathscr{N}} = 2
  higher derivative supergravity}},  {\em JHEP} {\bf 02} (2023) 110,
  [\href{http://arxiv.org/abs/2212.10590}{{\tt arXiv:2212.10590}}].

\bibitem{Hristov:2024cgj}
K.~Hristov, {\it {Equivariant localization and gluing rules in 4d
  $\mathcal{N}=2$ higher derivative supergravity}},  6, 2024.
\newblock \href{http://arxiv.org/abs/2406.18648}{{\tt arXiv:2406.18648}}.

\bibitem{Andrianopoli:1996cm}
L.~Andrianopoli, M.~Bertolini, A.~Ceresole, R.~D'Auria, S.~Ferrara, P.~Fre, and
  T.~Magri, {\it {N=2 supergravity and N=2 superYang-Mills theory on general
  scalar manifolds: Symplectic covariance, gaugings and the momentum map}},
  {\em J. Geom. Phys.} {\bf 23} (1997) 111--189,
  [\href{http://arxiv.org/abs/hep-th/9605032}{{\tt hep-th/9605032}}].

\bibitem{Lauria:2020rhc}
E.~Lauria and A.~Van~Proeyen, {\em {${\cal N}=2$ Supergravity in $D=4,5,6$
  Dimensions}}, vol.~966.
\newblock 3, 2020.

\bibitem{Couzens:2018wnk}
C.~Couzens, J.~P. Gauntlett, D.~Martelli, and J.~Sparks, {\it {A geometric dual
  of $c$-extremization}},  {\em JHEP} {\bf 01} (2019) 212,
  [\href{http://arxiv.org/abs/1810.11026}{{\tt arXiv:1810.11026}}].

\bibitem{Gauntlett:2018dpc}
J.~P. Gauntlett, D.~Martelli, and J.~Sparks, {\it {Toric geometry and the dual
  of $c$-extremization}},  {\em JHEP} {\bf 01} (2019) 204,
  [\href{http://arxiv.org/abs/1812.05597}{{\tt arXiv:1812.05597}}].

\bibitem{Hosseini:2019use}
S.~M. Hosseini and A.~Zaffaroni, {\it {Proving the equivalence of
  $c$-extremization and its gravitational dual for all toric quivers}},  {\em
  JHEP} {\bf 03} (2019) 108, [\href{http://arxiv.org/abs/1901.05977}{{\tt
  arXiv:1901.05977}}].

\bibitem{Hosseini:2019ddy}
S.~M. Hosseini and A.~Zaffaroni, {\it {Geometry of $\mathcal{I}$-extremization
  and black holes microstates}},  {\em JHEP} {\bf 07} (2019) 174,
  [\href{http://arxiv.org/abs/1904.04269}{{\tt arXiv:1904.04269}}].

\bibitem{Gauntlett:2019roi}
J.~P. Gauntlett, D.~Martelli, and J.~Sparks, {\it {Toric geometry and the dual
  of ${\cal I}$-extremization}},  {\em JHEP} {\bf 06} (2019) 140,
  [\href{http://arxiv.org/abs/1904.04282}{{\tt arXiv:1904.04282}}].

\bibitem{Kim:2019umc}
H.~Kim and N.~Kim, {\it {Black holes with baryonic charge and
  $\mathcal{I}$-extremization}},  {\em JHEP} {\bf 11} (2019) 050,
  [\href{http://arxiv.org/abs/1904.05344}{{\tt arXiv:1904.05344}}].

\bibitem{Boido:2022mbe}
A.~Boido, J.~P. Gauntlett, D.~Martelli, and J.~Sparks, {\it {Gravitational
  Blocks, Spindles and GK Geometry}},  {\em Commun. Math. Phys.} {\bf 403}
  (2023), no.~2 917--1003, [\href{http://arxiv.org/abs/2211.02662}{{\tt
  arXiv:2211.02662}}].

\bibitem{Aganagic:2003db}
M.~Aganagic, A.~Klemm, M.~Marino, and C.~Vafa, {\it {The Topological vertex}},
  {\em Commun. Math. Phys.} {\bf 254} (2005) 425--478,
  [\href{http://arxiv.org/abs/hep-th/0305132}{{\tt hep-th/0305132}}].

\bibitem{Okounkov:2003sp}
A.~Okounkov, N.~Reshetikhin, and C.~Vafa, {\it {Quantum Calabi-Yau and
  classical crystals}},  {\em Prog. Math.} {\bf 244} (2006) 597,
  [\href{http://arxiv.org/abs/hep-th/0309208}{{\tt hep-th/0309208}}].

\bibitem{Iqbal:2007ii}
A.~Iqbal, C.~Kozcaz, and C.~Vafa, {\it {The Refined topological vertex}},  {\em
  JHEP} {\bf 10} (2009) 069, [\href{http://arxiv.org/abs/hep-th/0701156}{{\tt
  hep-th/0701156}}].

\bibitem{Huang:2010kf}
M.-x. Huang and A.~Klemm, {\it {Direct integration for general $\Omega$
  backgrounds}},  {\em Adv. Theor. Math. Phys.} {\bf 16} (2012), no.~3
  805--849, [\href{http://arxiv.org/abs/1009.1126}{{\tt arXiv:1009.1126}}].

\bibitem{Krefl:2010fm}
D.~Krefl and J.~Walcher, {\it {Extended Holomorphic Anomaly in Gauge Theory}},
  {\em Lett. Math. Phys.} {\bf 95} (2011) 67--88,
  [\href{http://arxiv.org/abs/1007.0263}{{\tt arXiv:1007.0263}}].

\bibitem{Nekrasov:2002qd}
N.~A. Nekrasov, {\it {Seiberg-Witten prepotential from instanton counting}},
  {\em Adv. Theor. Math. Phys.} {\bf 7} (2003), no.~5 831--864,
  [\href{http://arxiv.org/abs/hep-th/0206161}{{\tt hep-th/0206161}}].

\bibitem{Nekrasov:2003rj}
N.~Nekrasov and A.~Okounkov, {\it {Seiberg-Witten theory and random
  partitions}},  {\em Prog. Math.} {\bf 244} (2006) 525--596,
  [\href{http://arxiv.org/abs/hep-th/0306238}{{\tt hep-th/0306238}}].

\bibitem{Bhattacharyya:2012ye}
S.~Bhattacharyya, A.~Grassi, M.~Marino, and A.~Sen, {\it {A One-Loop Test of
  Quantum Supergravity}},  {\em Class. Quant. Grav.} {\bf 31} (2014) 015012,
  [\href{http://arxiv.org/abs/1210.6057}{{\tt arXiv:1210.6057}}].

\bibitem{Liu:2017vbl}
J.~T. Liu, L.~A. Pando~Zayas, V.~Rathee, and W.~Zhao, {\it {One-Loop Test of
  Quantum Black Holes in anti\textendash{}de Sitter Space}},  {\em Phys. Rev.
  Lett.} {\bf 120} (2018), no.~22 221602,
  [\href{http://arxiv.org/abs/1711.01076}{{\tt arXiv:1711.01076}}].

\bibitem{Hristov:2021zai}
K.~Hristov and V.~Reys, {\it {Factorization of log-corrections in
  AdS$_{4}$/CFT$_{3}$ from supergravity localization}},  {\em JHEP} {\bf 12}
  (2021) 031, [\href{http://arxiv.org/abs/2107.12398}{{\tt arXiv:2107.12398}}].

\bibitem{Bobev:2023dwx}
N.~Bobev, M.~David, J.~Hong, V.~Reys, and X.~Zhang, {\it {A compendium of
  logarithmic corrections in AdS/CFT}},  {\em JHEP} {\bf 04} (2024) 020,
  [\href{http://arxiv.org/abs/2312.08909}{{\tt arXiv:2312.08909}}].

\bibitem{Gautason:2023igo}
F.~F. Gautason, V.~G.~M. Puletti, and J.~van Muiden, {\it {Quantized strings
  and instantons in holography}},  {\em JHEP} {\bf 08} (2023) 218,
  [\href{http://arxiv.org/abs/2304.12340}{{\tt arXiv:2304.12340}}].

\bibitem{Beccaria:2023ujc}
M.~Beccaria, S.~Giombi, and A.~A. Tseytlin, {\it {Instanton contributions to
  the ABJM free energy from quantum M2 branes}},  {\em JHEP} {\bf 10} (2023)
  029, [\href{http://arxiv.org/abs/2307.14112}{{\tt arXiv:2307.14112}}].

\bibitem{Beccaria:2023sph}
M.~Beccaria, S.~Giombi, and A.~A. Tseytlin, {\it {(2,0) theory on S5
  \texttimes{} S1 and quantum M2 branes}},  {\em Nucl. Phys. B} {\bf 998}
  (2024) 116400, [\href{http://arxiv.org/abs/2309.10786}{{\tt
  arXiv:2309.10786}}].

\bibitem{Kim:2009wb}
S.~Kim, {\it {The Complete superconformal index for N=6 Chern-Simons theory}},
  {\em Nucl. Phys. B} {\bf 821} (2009) 241--284,
  [\href{http://arxiv.org/abs/0903.4172}{{\tt arXiv:0903.4172}}]. [Erratum:
  Nucl.Phys.B 864, 884 (2012)].

\bibitem{Kapustin:2011jm}
A.~Kapustin and B.~Willett, {\it {Generalized Superconformal Index for Three
  Dimensional Field Theories}},  \href{http://arxiv.org/abs/1106.2484}{{\tt
  arXiv:1106.2484}}.

\bibitem{Beem:2012mb}
C.~Beem, T.~Dimofte, and S.~Pasquetti, {\it {Holomorphic Blocks in Three
  Dimensions}},  {\em JHEP} {\bf 12} (2014) 177,
  [\href{http://arxiv.org/abs/1211.1986}{{\tt arXiv:1211.1986}}].

\bibitem{Benini:2015noa}
F.~Benini and A.~Zaffaroni, {\it {A topologically twisted index for
  three-dimensional supersymmetric theories}},  {\em JHEP} {\bf 07} (2015) 127,
  [\href{http://arxiv.org/abs/1504.03698}{{\tt arXiv:1504.03698}}].

\bibitem{Benini:2015eyy}
F.~Benini, K.~Hristov, and A.~Zaffaroni, {\it {Black hole microstates in
  AdS$_{4}$ from supersymmetric localization}},  {\em JHEP} {\bf 05} (2016)
  054, [\href{http://arxiv.org/abs/1511.04085}{{\tt arXiv:1511.04085}}].

\bibitem{Hosseini:2022vho}
S.~M. Hosseini and A.~Zaffaroni, {\it {The large N limit of topologically
  twisted indices: a direct approach}},  {\em JHEP} {\bf 12} (2022) 025,
  [\href{http://arxiv.org/abs/2209.09274}{{\tt arXiv:2209.09274}}].

\bibitem{Bobev:2023lkx}
N.~Bobev, J.~Hong, and V.~Reys, {\it {Large N partition functions of 3d
  holographic SCFTs}},  {\em JHEP} {\bf 08} (2023) 119,
  [\href{http://arxiv.org/abs/2304.01734}{{\tt arXiv:2304.01734}}].

\bibitem{Bobev:2024mqw}
N.~Bobev, S.~Choi, J.~Hong, and V.~Reys, {\it {Superconformal indices of 3d $
  \mathcal{N} $ = 2 SCFTs and holography}},  {\em JHEP} {\bf 10} (2024) 121,
  [\href{http://arxiv.org/abs/2407.13177}{{\tt arXiv:2407.13177}}].

\bibitem{Inglese:2023wky}
M.~Inglese, D.~Martelli, and A.~Pittelli, {\it {The spindle index from
  localization}},  {\em J. Phys. A} {\bf 57} (2024), no.~8 085401,
  [\href{http://arxiv.org/abs/2303.14199}{{\tt arXiv:2303.14199}}].

\bibitem{Colombo:2024mts}
E.~Colombo, S.~M. Hosseini, D.~Martelli, A.~Pittelli, and A.~Zaffaroni, {\it
  {Microstates of Accelerating and Supersymmetric AdS4 Black Holes from the
  Spindle Index}},  {\em Phys. Rev. Lett.} {\bf 133} (2024), no.~3 031603,
  [\href{http://arxiv.org/abs/2404.07173}{{\tt arXiv:2404.07173}}].

\bibitem{Hosseini:2025mgf}
S.~M. Hosseini and A.~Zaffaroni, {\it {$ \mathcal{I} $-extremization for
  AdS$_{4}$ black holes: master volume, free energy, and baryonic charges}},
  {\em JHEP} {\bf 08} (2025) 100, [\href{http://arxiv.org/abs/2505.10626}{{\tt
  arXiv:2505.10626}}].

\bibitem{Plebanski:1976gy}
J.~F. Plebanski and M.~Demianski, {\it {Rotating, charged, and uniformly
  accelerating mass in general relativity}},  {\em Annals Phys.} {\bf 98}
  (1976) 98--127.

\bibitem{Alonso-Alberca:2000zeh}
N.~Alonso-Alberca, P.~Meessen, and T.~Ortin, {\it {Supersymmetry of topological
  Kerr-Newman-Taub-NUT-AdS space-times}},  {\em Class. Quant. Grav.} {\bf 17}
  (2000) 2783--2798, [\href{http://arxiv.org/abs/hep-th/0003071}{{\tt
  hep-th/0003071}}].

\bibitem{Katz:1996fh}
S.~H. Katz, A.~Klemm, and C.~Vafa, {\it {Geometric engineering of quantum field
  theories}},  {\em Nucl. Phys. B} {\bf 497} (1997) 173--195,
  [\href{http://arxiv.org/abs/hep-th/9609239}{{\tt hep-th/9609239}}].

\bibitem{Alexandrov:2023wdj}
S.~Alexandrov, M.~Mari\~no, and B.~Pioline, {\it {Resurgence of Refined
  Topological Strings and Dual Partition Functions}},  {\em SIGMA} {\bf 20}
  (2024) 073, [\href{http://arxiv.org/abs/2311.17638}{{\tt arXiv:2311.17638}}].

\bibitem{Hosseini:2019iad}
S.~M. Hosseini, K.~Hristov, and A.~Zaffaroni, {\it {Gluing gravitational blocks
  for AdS black holes}},  {\em JHEP} {\bf 12} (2019) 168,
  [\href{http://arxiv.org/abs/1909.10550}{{\tt arXiv:1909.10550}}].

\bibitem{BenettiGenolini:2023ndb}
P.~Benetti~Genolini, J.~P. Gauntlett, and J.~Sparks, {\it {Equivariant
  localization for AdS/CFT}},  {\em JHEP} {\bf 02} (2024) 015,
  [\href{http://arxiv.org/abs/2308.11701}{{\tt arXiv:2308.11701}}].

\bibitem{BenettiGenolini:2024lbj}
P.~Benetti~Genolini, J.~P. Gauntlett, Y.~Jiao, A.~L{\"u}scher, and J.~Sparks,
  {\it {Equivariant localization for D = 4 gauged supergravity}},  {\em JHEP}
  {\bf 08} (2025) 211, [\href{http://arxiv.org/abs/2412.07828}{{\tt
  arXiv:2412.07828}}].

\bibitem{deWit:2017cle}
B.~de~Wit and V.~Reys, {\it {Euclidean supergravity}},  {\em JHEP} {\bf 12}
  (2017) 011, [\href{http://arxiv.org/abs/1706.04973}{{\tt arXiv:1706.04973}}].

\bibitem{Aharony:2008gk}
O.~Aharony, O.~Bergman, and D.~L. Jafferis, {\it {Fractional M2-branes}},  {\em
  JHEP} {\bf 11} (2008) 043, [\href{http://arxiv.org/abs/0807.4924}{{\tt
  arXiv:0807.4924}}].

\bibitem{Bergman:2009zh}
O.~Bergman and S.~Hirano, {\it {Anomalous radius shift in AdS(4)/CFT(3)}},
  {\em JHEP} {\bf 07} (2009) 016, [\href{http://arxiv.org/abs/0902.1743}{{\tt
  arXiv:0902.1743}}].

\bibitem{Matsumoto:2013nya}
S.~Matsumoto and S.~Moriyama, {\it {ABJ Fractional Brane from ABJM Wilson
  Loop}},  {\em JHEP} {\bf 03} (2014) 079,
  [\href{http://arxiv.org/abs/1310.8051}{{\tt arXiv:1310.8051}}].

\bibitem{Honda:2014npa}
M.~Honda and K.~Okuyama, {\it {Exact results on ABJ theory and the refined
  topological string}},  {\em JHEP} {\bf 08} (2014) 148,
  [\href{http://arxiv.org/abs/1405.3653}{{\tt arXiv:1405.3653}}].

\bibitem{Atiyah:1978ri}
M.~F. Atiyah, N.~J. Hitchin, V.~G. Drinfeld, and Y.~I. Manin, {\it
  {Construction of Instantons}},  {\em Phys. Lett. A} {\bf 65} (1978) 185--187.

\bibitem{deBoer:1996mp}
J.~de~Boer, K.~Hori, H.~Ooguri, and Y.~Oz, {\it {Mirror symmetry in
  three-dimensional gauge theories, quivers and D-branes}},  {\em Nucl. Phys.
  B} {\bf 493} (1997) 101--147,
  [\href{http://arxiv.org/abs/hep-th/9611063}{{\tt hep-th/9611063}}].

\bibitem{Porrati:1996xi}
M.~Porrati and A.~Zaffaroni, {\it {M theory origin of mirror symmetry in
  three-dimensional gauge theories}},  {\em Nucl. Phys. B} {\bf 490} (1997)
  107--120, [\href{http://arxiv.org/abs/hep-th/9611201}{{\tt hep-th/9611201}}].

\bibitem{PeterBKronheimer:1990zmj}
P.~B. Kronheimer and H.~Nakajima, {\it {Yang-Mills instantons on ALE
  gravitational instantons}},  {\em Math. Ann.} {\bf 288} (1990), no.~1
  263--307.

\bibitem{Imamura:2008nn}
Y.~Imamura and K.~Kimura, {\it {On the moduli space of elliptic
  Maxwell-Chern-Simons theories}},  {\em Prog. Theor. Phys.} {\bf 120} (2008)
  509--523, [\href{http://arxiv.org/abs/0806.3727}{{\tt arXiv:0806.3727}}].

\bibitem{Cremonesi:2016nbo}
S.~Cremonesi, N.~Mekareeya, and A.~Zaffaroni, {\it {The moduli spaces of 3d $
  \mathcal{N}\ge 2 $ Chern-Simons gauge theories and their Hilbert series}},
  {\em JHEP} {\bf 10} (2016) 046, [\href{http://arxiv.org/abs/1607.05728}{{\tt
  arXiv:1607.05728}}].

\bibitem{Cacciatori:2009iz}
S.~L. Cacciatori and D.~Klemm, {\it {Supersymmetric AdS(4) black holes and
  attractors}},  {\em JHEP} {\bf 01} (2010) 085,
  [\href{http://arxiv.org/abs/0911.4926}{{\tt arXiv:0911.4926}}].

\bibitem{DallAgata:2010ejj}
G.~Dall'Agata and A.~Gnecchi, {\it {Flow equations and attractors for black
  holes in N = 2 U(1) gauged supergravity}},  {\em JHEP} {\bf 03} (2011) 037,
  [\href{http://arxiv.org/abs/1012.3756}{{\tt arXiv:1012.3756}}].

\bibitem{Hristov:2010ri}
K.~Hristov and S.~Vandoren, {\it {Static supersymmetric black holes in
  AdS$_{4}$ with spherical symmetry}},  {\em JHEP} {\bf 04} (2011) 047,
  [\href{http://arxiv.org/abs/1012.4314}{{\tt arXiv:1012.4314}}].

\bibitem{Halmagyi:2013sla}
N.~Halmagyi, M.~Petrini, and A.~Zaffaroni, {\it {BPS black holes in $AdS_{4}$
  from M-theory}},  {\em JHEP} {\bf 08} (2013) 124,
  [\href{http://arxiv.org/abs/1305.0730}{{\tt arXiv:1305.0730}}].

\bibitem{Hristov:2018spe}
K.~Hristov, S.~Katmadas, and C.~Toldo, {\it {Rotating attractors and BPS black
  holes in $AdS_4$}},  {\em JHEP} {\bf 01} (2019) 199,
  [\href{http://arxiv.org/abs/1811.00292}{{\tt arXiv:1811.00292}}].

\bibitem{Hristov:2019mqp}
K.~Hristov, S.~Katmadas, and C.~Toldo, {\it {Matter-coupled supersymmetric
  Kerr-Newman-AdS$_4$ black holes}},  {\em Phys. Rev. D} {\bf 100} (2019),
  no.~6 066016, [\href{http://arxiv.org/abs/1907.05192}{{\tt
  arXiv:1907.05192}}].

\bibitem{Ferrero:2020twa}
P.~Ferrero, J.~P. Gauntlett, J.~M.~P. Ipi\~na, D.~Martelli, and J.~Sparks, {\it
  {Accelerating black holes and spinning spindles}},  {\em Phys. Rev. D} {\bf
  104} (2021), no.~4 046007, [\href{http://arxiv.org/abs/2012.08530}{{\tt
  arXiv:2012.08530}}].

\bibitem{Couzens:2021cpk}
C.~Couzens, {\it {A tale of (M)2 twists}},  {\em JHEP} {\bf 03} (2022) 078,
  [\href{http://arxiv.org/abs/2112.04462}{{\tt arXiv:2112.04462}}].

\bibitem{Ferrero:2021etw}
P.~Ferrero, J.~P. Gauntlett, and J.~Sparks, {\it {Supersymmetric spindles}},
  {\em JHEP} {\bf 01} (2022) 102, [\href{http://arxiv.org/abs/2112.01543}{{\tt
  arXiv:2112.01543}}].

\bibitem{Hristov:2023rel}
K.~Hristov and M.~Suh, {\it {Spindle black holes in AdS$_{4} \times$
  SE$_{7}$}},  {\em JHEP} {\bf 10} (2023) 141,
  [\href{http://arxiv.org/abs/2307.10378}{{\tt arXiv:2307.10378}}].

\bibitem{Hosseini:2016tor}
S.~M. Hosseini and A.~Zaffaroni, {\it {Large $N$ matrix models for 3d ${\cal
  N}=2$ theories: twisted index, free energy and black holes}},  {\em JHEP}
  {\bf 08} (2016) 064, [\href{http://arxiv.org/abs/1604.03122}{{\tt
  arXiv:1604.03122}}].

\bibitem{Hosseini:2016ume}
S.~M. Hosseini and N.~Mekareeya, {\it {Large $N$ topologically twisted index:
  necklace quivers, dualities, and Sasaki-Einstein spaces}},  {\em JHEP} {\bf
  08} (2016) 089, [\href{http://arxiv.org/abs/1604.03397}{{\tt
  arXiv:1604.03397}}].

\bibitem{Hosseini:2021fge}
S.~M. Hosseini, K.~Hristov, and A.~Zaffaroni, {\it {Rotating multi-charge
  spindles and their microstates}},  {\em JHEP} {\bf 07} (2021) 182,
  [\href{http://arxiv.org/abs/2104.11249}{{\tt arXiv:2104.11249}}].

\bibitem{Faedo:2021nub}
F.~Faedo and D.~Martelli, {\it {D4-branes wrapped on a spindle}},  {\em JHEP}
  {\bf 02} (2022) 101, [\href{http://arxiv.org/abs/2111.13660}{{\tt
  arXiv:2111.13660}}].

\bibitem{Martelli:2005tp}
D.~Martelli, J.~Sparks, and S.-T. Yau, {\it {The Geometric dual of
  a-maximisation for Toric Sasaki-Einstein manifolds}},  {\em Commun. Math.
  Phys.} {\bf 268} (2006) 39--65,
  [\href{http://arxiv.org/abs/hep-th/0503183}{{\tt hep-th/0503183}}].

\bibitem{Butti:2005vn}
A.~Butti and A.~Zaffaroni, {\it {R-charges from toric diagrams and the
  equivalence of a-maximization and Z-minimization}},  {\em JHEP} {\bf 11}
  (2005) 019, [\href{http://arxiv.org/abs/hep-th/0506232}{{\tt
  hep-th/0506232}}].

\bibitem{Butti:2005ps}
A.~Butti and A.~Zaffaroni, {\it {From toric geometry to quiver gauge theory:
  The Equivalence of a-maximization and Z-minimization}},  {\em Fortsch. Phys.}
  {\bf 54} (2006) 309--316, [\href{http://arxiv.org/abs/hep-th/0512240}{{\tt
  hep-th/0512240}}].

\bibitem{Martelli:2006yb}
D.~Martelli, J.~Sparks, and S.-T. Yau, {\it {Sasaki-Einstein manifolds and
  volume minimisation}},  {\em Commun. Math. Phys.} {\bf 280} (2008) 611--673,
  [\href{http://arxiv.org/abs/hep-th/0603021}{{\tt hep-th/0603021}}].

\bibitem{Amariti:2011uw}
A.~Amariti, C.~Klare, and M.~Siani, {\it {The Large N Limit of Toric
  Chern-Simons Matter Theories and Their Duals}},  {\em JHEP} {\bf 10} (2012)
  019, [\href{http://arxiv.org/abs/1111.1723}{{\tt arXiv:1111.1723}}].

\bibitem{Gauntlett:2019pqg}
J.~P. Gauntlett, D.~Martelli, and J.~Sparks, {\it {Fibred GK geometry and
  supersymmetric $AdS$ solutions}},  {\em JHEP} {\bf 11} (2019) 176,
  [\href{http://arxiv.org/abs/1910.08078}{{\tt arXiv:1910.08078}}].

\bibitem{Papadimitriou:2017kzw}
I.~Papadimitriou, {\it {Supercurrent anomalies in 4d SCFTs}},  {\em JHEP} {\bf
  07} (2017) 038, [\href{http://arxiv.org/abs/1703.04299}{{\tt
  arXiv:1703.04299}}].

\bibitem{Intriligator:2003jj}
K.~A. Intriligator and B.~Wecht, {\it {The Exact superconformal R symmetry
  maximizes a}},  {\em Nucl. Phys. B} {\bf 667} (2003) 183--200,
  [\href{http://arxiv.org/abs/hep-th/0304128}{{\tt hep-th/0304128}}].

\bibitem{Hanada:2012si}
M.~Hanada, M.~Honda, Y.~Honma, J.~Nishimura, S.~Shiba, and Y.~Yoshida, {\it
  {Numerical studies of the ABJM theory for arbitrary N at arbitrary coupling
  constant}},  {\em JHEP} {\bf 05} (2012) 121,
  [\href{http://arxiv.org/abs/1202.5300}{{\tt arXiv:1202.5300}}].

\bibitem{Dedushenko:2014nya}
M.~Dedushenko and E.~Witten, {\it {Some Details On The Gopakumar-Vafa and
  Ooguri-Vafa Formulas}},  {\em Adv. Theor. Math. Phys.} {\bf 20} (2016)
  1--133, [\href{http://arxiv.org/abs/1411.7108}{{\tt arXiv:1411.7108}}].

\bibitem{Grassi:2014zfa}
A.~Grassi, Y.~Hatsuda, and M.~Marino, {\it {Topological Strings from Quantum
  Mechanics}},  {\em Annales Henri Poincare} {\bf 17} (2016), no.~11
  3177--3235, [\href{http://arxiv.org/abs/1410.3382}{{\tt arXiv:1410.3382}}].

\bibitem{Marino:2015nla}
M.~Marino, {\it {Spectral theory and mirror symmetry.}},  {\em Proc. Symp. Pure
  Math.} {\bf 98} (2018) 259, [\href{http://arxiv.org/abs/1506.07757}{{\tt
  arXiv:1506.07757}}].

\bibitem{Codesido:2015dia}
S.~Codesido, A.~Grassi, and M.~Marino, {\it {Spectral Theory and Mirror Curves
  of Higher Genus}},  {\em Annales Henri Poincare} {\bf 18} (2017), no.~2
  559--622, [\href{http://arxiv.org/abs/1507.02096}{{\tt arXiv:1507.02096}}].

\end{thebibliography}\endgroup
%%%%%%%%%%%%%%%%%%%%%%%%%%%%%%%%%%%%%

\end{document}